\documentclass[a4paper,11pt]{article}
\usepackage{aaskaiid2}
\usepackage{wasysym}
\usepackage{pjournal}
\usepackage{graphicx}
\usepackage{comment}
\usepackage{orcidlink}

\newcommand{\dnu}{\Delta\nu}
\newcommand{\kpar}{k_{\parallel}}

\title{Observations of the Cosmic Dawn and Epoch of Reionization with the SKAO: Observational Lessons Learned from Precursors and Pathfinder Instruments}
\ShortTitle{SKAO CD/EOR: observational lessons}

\author[1,2]{Eloy de Lera Acedo\orcidlink{0000-0001-8530-6989}}
\author[3]{James Aguirre}
\author[1,2]{Dominic Anstey\orcidlink{0000-0003-1742-7417}}
\author[4,5]{Nichole Barry\orcidlink{0000-0003-2064-6979}}
\author[6,7,8]{Gianni Bernardi*\orcidlink{0000-0002-0916-7443}}
\author[9]{Somnath Bharadwaj}
\author[10,11]{Anthony Brown \orcidlink{0000-0002-7033-7544}}
\author[12]{Jean Cavillot\orcidlink{0000-0001-5731-1721}}
\author[13]{Suman Chatterjee}
\author[14]{Samir Choudhuri}
\author[15]{Tyler Cox \orcidlink{0009-0008-2574-3878}}
\author[1,2]{John Cumner \orcidlink{0000-0001-8479-3746}}
\author[16]{Abhirup Datta}
\author[1]{Fred Dulwich}
\author[14]{Khandakar Md Asif Elahi~\orcidlink{0000-0003-1206-8689}}
\author[1]{Andrew Faulkner}
\author[9]{Sukhdeep Singh Gill}
\author[1,2]{Quentin Gueuning\orcidlink{0000-0002-0127-6784}}
\author[17]{Daniel C. Jacobs}
\author[18]{Nicholas Kern\orcidlink{0000-0002-8825-1669}}
\author[7,13]{Piyanat Kittiwisit\orcidlink{0000-0003-0953-313X}}
\author[1,2]{Yuchen Liu}
\author[3]{Zachary Martinot}
\author[1,2]{Ashish Mhaske}
\author[19,20]{Florent Mertens\orcidlink{0000-0003-3802-4289}}
\author[18,21]{Vincent McKay}
\author[20]{Satyapan Munshi}
\author[17,22]{Steven Murray\orcidlink{0000-0003-3059-3823}}
\author[23,24]{Chuneeta D. Nunhokee}
\author[1,2]{Oscar Sage David O'Hara\orcidlink{0009-0006-3633-5816}}
\author[16]{Samit K. Pal\orcidlink{0000-0002-2271-4165}}
\author[28,29]{Robert Pascua}
\author[16]{Rashmi Sagar}  
\author[14]{Shouvik Sarkar}
\author[25]{Shiv Sethi}
\author[26]{Sarod Yatawatta\orcidlink{0000-0001-5619-4017}}
\author[1,2,27]{Oskar Zetterstrom\orcidlink{0000-0002-5338-1789}}

\affiliation[1]{Cavendish Astrophysics, University of Cambridge, Cambridge CB3 0HE, UK}
\affiliation[2]{Kavli Institute for Cosmology in Cambridge, University of Cambridge, Cambridge CB3 0HA, UK}
\affiliation[3]{Department of Physics and Astronomy, University of Pennsylvania, 209 South 33rd Street, Philadelphia, PA 19104, USA}
\affiliation[4]{School of Physics, The University of New South Wales, Australia}
\affiliation[5]{ARC Centre of Excellence for All Sky Astrophysics in 3 Dimensions (ASTRO 3D), Australia}
\affiliation[6]{INAF–Istituto di Radio Astronomia (IRA), Via Piero Gobetti 101, I-40129 Bologna, Italy}
\affiliation[7]{South African Radio Astronomy Observatory, Cape Town 7700, South Africa}
\affiliation[8]{Centre for Radio Astronomy Techniques and Technologies (RATT), Department of Physics and Electronics, Rhodes University, Makhanda 6140, South Africa}
\affiliation[9]{Department of Physics, Indian Institute of Technology Kharagpur, Kharagpur - 721 302, India}
\affiliation[10]{School of Electronic Engineering and Computer Science, Queen Mary, University of London, London E1 4NS, UK}
\affiliation[11]{Department of Electrical and Electronic Engineering, The University of Manchester, Manchester M13 9PL, UK}
\affiliation[12]{Institute of Information and Communication Technologies, Electronics and Applied Mathematics (ICTEAM), Universite Catholique}
\affiliation[13]{Department of Physics and Astronomy, University of the Western Cape, 7535 Bellville, Cape Town, South Africa}
\affiliation[14]{Centre for Strings, Gravitation and Cosmology, Department of Physics, Indian Institute of Technology Madras, Chennai 600036, India}
\affiliation[15]{Department of Astronomy, University of California, Berkeley, CA 94720, USA}
\affiliation[16]{Department of Astronomy, Astrophysics and Space Engineering, Indian Institute of Technology Indore, Indore 453552, India}
\affiliation[17]{School of Earth and Space Exploration, Arizona State University, Tempe,
AZ 85287, USA}
\affiliation[18]{Department of Physics, University of Michigan, Ann Arbor, MI 48109}
\affiliation[19]{LUX, Observatoire de Paris, PSL Research University, CNRS, Sorbonne Université, F-75014 Paris, France}
\affiliation[20]{Kapteyn Astronomical Institute, University of Groningen, PO Box 800, 9700 AV Groningen, The Netherlands}
\affiliation[21]{MIT Kavli Institute, Massachusetts Institute of Technology, Cambridge, MA 02139, USA}
\affiliation[22]{Scuola Normale Superiore, I-56126 Pisa PI, Italy}
\affiliation[23]{International Centre for Radio Astronomy Research, Curtin University, Bentley, WA, Australia}
\affiliation[24]{ARC. Centre of Excellence for All Sky Astrophysics in 3 Dimensions (ASTRO 3D), Bentley, Australia}
\affiliation[25]{Raman Research Institute, C. V. Raman Avenue, Sadashivanagar, Bengaluru 560080, India}
\affiliation[26]{Astron, PO Box 2, 7990 AA Dwingeloo, The Netherlands}
\affiliation[27]{Division of Electromagnetic Engineering and Fusion Science, KTH Royal Institute of Technology, Stockholm 11428, Sweden}
\affiliation[28]{Dunlap Institute for Astronomy and Astrophysics, University of Toronto, Toronto, ON M5S 3H4, Canada}
\affiliation[29]{Perimeter Institute for Theoretical Physics, Waterloo, ON N2L 2YF, Canada}

\emailAdd{gianni.bernardi@inaf.it}

\abstract{This chapter summarizes the observational lessons learned after two decades of observations of the Cosmic Dawn (CD) and Epoch of Reionization (EoR) with SKAO pathfinders and precursors. We will describe the effort towards building accurate simulation pipelines for actual observations and summarize the approaches that different groups have taken to calibrate and mitigate systematic effects such as sky model incompleteness, limited instrument models and antenna mutual coupling. We conclude by discussing the impact that these lessons may have on the design and analysis of upcoming SKAO observations of the Cosmic Dawn and Epoch of Reionization.}

\begin{document}
\maketitle

\section{Introduction}

The detection of the redshifted 21~cm line has always been recognized as challenging due to the presence of foreground emission from galactic and extra-galactic sources that is orders of magnitude brighter than the underlying cosmological signal of interest. This sets up a delicate signal separation problem, which is complicated by the fact that the 21\,cm cosmological signal, the low-frequency foreground sky, and the instrumental response of the telescope are not known a priori by the observer to high precision.
In this chapter we will summarize a decade of 21 cm observations carried out by the SKAO precursors/pathfinders, i.e. the upgraded Giant Metrewave Radio Telescope (uGMRT), the Hydrogen Epoch of Reioniozation Array (HERA), the Low Frequency Array (LOFAR) and the Murchison Widefield Array (MWA). We will discuss the specific challenges that the various telescopes faced and how they addressed them, drawing lessons for future SKAO observations - a topic covered in further detail in another chapter of this book \citep{Trott01.2026.SKA}. The chapter is organized as follows: in Section~\ref{sec:simulations_validation} we describe the development of end-to-end simulation pipelines necessary to validate 21 cm observations; the lessons learned from pathfinders and precursors are described in Section~\ref{sec:pathfinders_precursors}; the implication of the SKA-Low antenna design for 21~cm observations and of mutual coupling among closely packed antennas is discussed in Sections~\ref{sec:skala} and \ref{sec:mutual}, respectively. Conclusions are presented in Section~\ref{sec:conclusions}.

\section{End-to-end simulations and pipeline validation}\label{sec:simulations_validation}

Many radio interferometers have taken enough data such that a detection of the 21-cm EoR signal could be made given the theoretical thermal noise floor; however, a definitive measurement has yet to be made.
This is rooted in the complex interplay between bright foregrounds \citep[a topic covered in further detail in another chapter of this book;][]{Burba01.2026.SKA} and the faint cosmological signal, including the modulation of foregrounds by the instrumental response, terrestrial systematics, and pipeline analyses of insufficient precision and accuracy.
To overcome this, teams have developed increasingly sophisticated pipelines to model and remove these  artifacts. Often, these are lossy and can bias the results \citep{LiuShaw2020}.
To ensure the fidelity of these pipelines, end-to-end validation approaches have played an increasingly important role due to the need to faithfully simulate low-level systematics and test our analysis assumptions (i.e., \citealt{Barry2019, Aguirre_HERA_e2eSims_2022, mertens2020}).

The core philosophy of an end-to-end pipeline validation is specific testing of a pipeline ability to handle a \emph{known} systematic, with the primary goal of verifying that an analysis pipeline recovers an unbiased estimate of the 21\,cm signal.
The general approach is to begin with a simulated 21\,cm signal, add a model of the foreground sky to it, and propagate their sum forward through an instrument model that includes as many of the known effects as possible, with sufficient realism to test the analysis pipeline. 
This is followed by pushing the mock data through the same (or as similar as possible) analysis pipeline as the real data, and comparing the resulting power spectrum with that of the input 21\,cm signal. 
This methodology is fundamentally different than trying to discover systematics, for example, via testing similar sets with various analyses to uncover an unknown systematic (e.g., jackknife testing, \citealt{beardsley_first_2016, HERA2022a}).

In comparison to more commonly used signal-injection tests, by which the mock data is formed by injecting simulated and boosted 21~cm signals to the real data, the end-to-end validation has a few important advantages.
First, it is possible to have many realizations of the mock data as noise and systematics can be independently controlled and are not tied to the data, allowing for more statistically robust testing.
Second, for power spectrum analysis that relies on inverse-covariance weighting, mismatch between the true data covariance and the covariance used in the analysis can lead to unaccounted signal loss \citep{cheng_characterizing_2018}. End-to-end validation completely avoids this situation as the covariance of the simulated mock data is known exactly.
Its biggest drawback, however, is that the simulation might not have systematics that the real data has, as one can only model and simulate known systematics.



A number of software packages has been developed by teams within the EoR community to enable end-to-end validation, each with different approaches and different assumptions.
At their core, these codes evaluate the radio interferometric measurement equation \citep[e.g.][]{Hamaker1996, Smirnov2011} with differing levels of accuracy, speed, and assumptions of astrophysical, terrestrial, or instrumental systematic effects.
Different 21\,cm telescopes, depending on their configuration and design, can make different approximations when it comes to simulating their data.
Wide-field, low-angular resolution telescopes like HERA, for example, can use a coarser sky model than high-angular resolution telescopes like the MWA or LOFAR; however, at the same time, compact array configurations like HERA will be more sensitive to diffuse Galactic emission. 
These differing requirements, along with recent advances in high-performance computing, have resulted in several visibility simulators and validation pipelines, and their inclusion as a robust foundation to the upper limits to the EoR 21-cm signal is now considered fairly standard.




\subsection{Visibility simulations}

Visibility simulators are required to forward-model sky-based signals through the instrument and into visibilities, which can then be used to validate pipelines. Accuracy and precision requirements dictate the style of simulator. On the ultra-precise side, \texttt{DP3} routines \citep{dijkema_dp3_2023}, \texttt{WODEN} \citep{line_woden_2022}, and \texttt{pyuvsim} \citep{lanman_pyuvsim_2019} enable high-fidelity simulations for the majority of precursors. These can be quite costly, and thus these simulators result in the need for dedicated software developers or rare usage.
Other simulators, like \texttt{matvis} \citep{Kittiwisit_matvis_2024}, \texttt{RIMEz}, \texttt{fftvis} \citep{Cox_fftvis_2025} and \texttt{BayesLIM} \citep{Kern2025} enable judicious approximations, often in concert with GPUs and other computational speed-ups, to accelerate the forward modeling process, and thus enable scaling-up to the large number of baselines measured by current-generation experiments.

Visibility simulators enable the forward-modeling of signals through the instrumental sensitivity response described by the radio interferometric measurement equation. In general, this is the bottleneck in ultra-precise simulations---the exact sensitivity response of each station, for each time, for each frequency, and for each point on the sky signal quickly becomes millions if not billions of operations. Modifications to the instrument response to test realistic variations for each station become even more costly. Station gain variations (i.e., \citealt{joseph_calibration_2020,chokshi_necessity_2024}) or mutual coupling \citep{ohara2025uncoupling} can be incorporated, but it is not something that can currently be done often. The foreground coupling to these instrumental station variations requires careful attention in the SKAO case (see Section~\ref{sec:mutual}).


Besides a robust and well-tested simulator, the first required component of an end-to-end simulation is the 21\,cm signal itself.
How accurate this simulation needs to be depends on the validation to be performed, and several approximations can be made that still allow for robust testing in many cases. 
For example, both the MWA (\citealt{barry_calibration_2016,byrne_fundamental_2019}) and HERA collaborations have used Gaussian random fields with a given input power spectrum as a proxy for the 21\,cm field  (\citealt{Aguirre_HERA_e2eSims_2022}, using \texttt{redshifted\_gaussian\_fields} \footnote{\url{https://github.com/zacharymartinot/redshifted_gaussian_fields}}). However,  21\,cm fluctuations during the EoR are intrinsically non-Gaussian,
so the Gaussian random field approximation does neglect some realism that may hide biases in the analysis pipeline.
Realistic patches of signal, tiled with boundary conditions, are the next level in simulation accuracy \citep{byrne_fundamental_2019}. However, this can produce statistical artifacts that affect the final result (i.e., \citealt{barry_improving_2019}). 
The most realistic approach is to forward-model near full-sky EoR simulations, but this comes at great computational cost and it has been only implemented by \cite{line_verifying_2024}, so far with a 50 square degree EoR model.
Even so, care must be taken when interpolating the coordinates of the underlying simulation to that of the curved sky.

Foregrounds are an important component to test the ability of a pipeline to handle extremely bright signals.
Generating a realistic model of the sky can be very computationally expensive, depending on the resolution and volume required. For example, extremely wide-field-of-view instruments like HERA and the MWA have close to 100,000 known unresolved sources at any given time. Radio catalogs at proper frequencies, like GLEAM \citep{hurley-walker_galactic_2017} and LoTSS \citep{shimwell_lofar_2017}, are commonly used as input. 
For compact and wide-field arrays like HERA, diffuse emission is the dominant foreground for the shortest baselines, but observations of diffuse emission with all-sky coverage at these frequencies do not exist.
While all-sky diffuse radio models are available \citep{Oliveira-Costa2008,Zheng2017}, they rely on principle component analysis of several incomplete maps at other frequencies. Combining them with known point sources can also tend to over-produce total sky emission, and the absolute temperature scale of the maps is somewhat uncertain \citep{Monsalve2021}.
The discretisation of diffuse sky models also poses questions about the use of optimal basis functions \citep{Lanman2022}: the \texttt{HEALPix} pixelisation is a common choice, but other bases such as spherical harmonics provide some advantages. 
The resolution of the required diffuse model also depends on the maximum simulated baseline of the experiment, with long baselines requiring extremely large numbers of pixels (e.g., $>10^7$ in \citealt{Kittiwisit_matvis_2024}). 





\subsection{Simulations of instrumental effects}

As the focus of 21\,cm experiments shifts from obtaining raw sensitivity to improving control over systematics, it becomes increasingly important to include known systematic effects in the mock data used for validation.
Many of the known systematics operate on the data at the visibility level, either through a multiplicative scaling (such as electronic gains or cable reflections) or an additive term (such as thermal noise or radio frequency interference, or RFI).
As such, this class of systematic effects may be included in validation analyses for relatively little computational cost, thereby enabling analysts to generate more realistic mock data in end-to-end tests (such as those in~\citealt{Aguirre_HERA_e2eSims_2022}).
Other systematic effects, such as element-to-element variations and some flavours of mutual coupling, instead modify the direction-dependent instrument response in a way that cannot be expressed as a linear operation on the idealized visibilities.
Consequently, this class of systematic effects is ill-suited for end-to-end validation tests due to their prohibitive computational cost, but simulations of these systematics may still be used in targeted pipeline tests.

The inclusion of known systematic effects, modeled following our knowledge of how the effects appear in the instrument, is a crucial aspect of validating 21\,cm data analyses.
Studies have shown that residual gains from an imperfect calibration (e.g, due to gain variations not captured by the calibration model or imperfect fits to cable reflections) can contaminate the 21\,cm signal by scattering foreground power into the EoR window~\citep[e.g.,][]{barry_calibration_2016,byrne_fundamental_2019,Kern2019}.
Alternatively, more aggressive mitigation strategies, such as the filtering strategy for mitigating mutual coupling in HERA observations \citep{Rath2025}, run the risk of attenuating the 21\,cm signal in the mitigation process \citep[e.g.,][]{Pascua:2025}.
Because residual systematics may contaminate the 21\,cm signal and our treatment of those systematics may bias our estimates of the 21\,cm signal if improperly accounted for, it is important for end-to-end validation tests to include as many realistic systematic effects as possible when generating mock data.

Although the community's understanding of systematic effects in 21\,cm data has shown continued growth over the years, some of them are not properly accounted for in realistic simulations (precluding tests of our analysis pipelines in a genuine end-to-end fashion).
Particularly important among these systematics is RFI, due to its prevalence and its ability to obscure the 21\,cm signal~\citep{Wilensky2020}.
Much of the bright RFI contaminates only a handful of channels (e.g., FM radio and ORBCOMM) and is therefore relatively easily handled with standard flagging procedures and inpainting algorithms, enabling analysts to instead focus on validating the inpainting algorithms rather than the flagging algorithms themselves~\citep{Aguirre_HERA_e2eSims_2022}.
Faint RFI that lies at or below the noise level, however, presents a much more challenging task for validation because it brings into question the accuracy of the flagging algorithm.
While there are some simulation-based tests of flagging algorithms in the literature~\citep[e.g.,][]{Offringa2012,Wilensky2020}, tests of RFI flagging fidelity have not been included in end-to-end validation tests due to the extremely complex and highly uncertain statistical distribution of faint RFI.

Polarization is another important systematic that has not yet been incorporated in simulations and validation pipelines as all existing efforts are focusing on obtaining the total intensity EoR power spectrum. However, \citet{Byrne2022} showed that polarized diffuse sky models are important for high-precision calibration.
As 21\,cm data become more sensitive, validation teams should seek to broaden the scope of systematic effects included in mock data, with a focus on known effects that are poorly understood (such as faint RFI and polarization).

\subsection{Pipeline Validation}
\label{sec_pipeline_validation}

Armed with an end-to-end mock dataset with a known input 21\,cm signal, 
the task is to perform relevant tests that verify the analysis algorithms and choices.
This can take two forms: (i) targeted tests of specific analysis steps, and 
(ii) end-to-end tests of signal recovery \citep{Aguirre_HERA_e2eSims_2022}. 

Targeted tests can be useful to identify key components of the analysis that introduce bias.
These tests often use limited subsets of data (and often without a full set of sky and instrumental components included), 
where specific data axes are prioritized (e.g., channel resolution and bandwidth to identify high-delay structures, or time sampling to identify temporal discontinuities). 
An example of the utility of this class of tests was encountered in the validation of Phase I HERA data, when a $\sim 10\%$ bias was discovered in the absolute calibration algorithm, and was corrected before the final data analysis \citep{HERA2022a}.


Targeted pipeline validation tests for direction-dependent (DD) calibration and Gaussian process regression (GPR) have been developed for LOFAR and NenuFAR 21-cm cosmology observations \citep{mevius2021,munshi2024first}, using a combination of forward simulations and signal injection tests. Both experiments follow a similar analysis: known sky sources are first subtracted from visibilities using a DD calibration and subtraction approach, after which residual unmodeled foregrounds are separated from the 21-cm signal in gridded visibilities using their spectral smoothness via GPR.

DD calibration involves a large number of free parameters, which increases the risk of overfitting and suppressing the 21-cm signal. The analysis of LOFAR EoR observations mitigates this risk by performing calibration only on baselines longer than a chosen cutoff, thereby confining any signal loss to that subset of baselines. Excluding these longer baselines during power spectrum estimation prevents the suppression from propagating into the final 21-cm signal measurement. Such a baseline cut, however, introduces excess noise on the shorter baselines \citep{Patil2016}. This has been studied extensively, both through analytical \citep{Millad2018} and numerical \citep{mevius2021} approaches. This excess noise can be partially mitigated by enforcing frequency smoothness constraints on the gain solutions, which reduces overfitting during calibration. 

GPR-based foreground subtraction approaches carry a risk of suppressing the 21-cm signal unless carefully validated. Analytical forms for the 21-cm signal covariance may not accurately represent the covariance of the true signal, resulting in signal loss \citep{Kern2021}. To address this, improved implementations such as ML-GPR \citep{Mertens2024} have been developed, which employ machine-learned covariance kernels trained on suites of 21-cm signal simulations, thus providing a more realistic prior. The robustness of GPR is assessed through signal injection tests on the gridded visibilities, performed over an ensemble of injected 21-cm signal amplitudes and power spectrum shapes. Recent extensions, such as cross-GPR \citep{munshi2025mitigating}, have been proposed to utilize the coherence of the 21-cm signal across nights, further improving the robustness of GPR to signal suppression.
Another extension includes the hierarchical Gaussian process \citep{diao2025}, where the model used allows for variation in kernel parameters between different lines of sight and strikes a balance between accuracy and robustness.


Another example of an end-to-end validation pipeline is FHD/$\epsilon$ppsilon \citep{Barry2019}. For every analysis done on data, a matching mock simulation of foregrounds and expected visibility noise is propagated through the pipeline all the way to power spectra for validation. However, performing full end-to-end validations for every data set, including realistic EoR signals and instrumental variation, is currently not computationally possible within any pipeline.

\section{Experience with pathfinders and precursors}
\label{sec:pathfinders_precursors}

\subsection{Overview of the LOFAR calibration architecture and lessons learned}
\label{sec:lofar_lessons}
First, we need to emphasize the particular characteristics of LOFAR EoR observations compared to any other LOFAR imaging observation. We observe, collect and curate data amounting to thousands of observing hours (peta-bytes of storage) needed for EoR science. A typical interferometric imaging observation on the other hand lasts only about 10 hours. Therefore, the requirements in precision and accuracy of these two observations are vastly different. Furthermore, LOFAR is an instrument serving many more diverse science goals other than the EoR. Therefore, we strive to solve all issues related to the data by adapting the data processing strategies rather than customizing the hardware.
In the following, we summarize the lessons learnt from research and data processing with LOFAR in the past decade \citep{NCP2013,Patil2017,mertens2020,mertens2025,Ceccotti2025}.

\begin{itemize}

\item The sky model plays a crucial role in calibration as well as foreground subtraction, and
its completeness is critical. Obviously, when we start processing a particular target in the sky, we do not have a sky model that is complete enough. It is possible to bootstrap this model with known sky catalogs, but such models are not complete for our purpose. Therefore,several iterations of calibration and imaging are necessary to build a sky model with sufficient completeness and accuracy. From a practical point of view, it is better to have an overly complex model and trim down with fine-tuning rather than the opposite, i.e., have a too simple model and expand.

\item The complexity of direction dependent calibration goes hand in hand with the complexity of the sky model. A sufficiently complete sky model will cover the full sky and the number of directions being calibrated should also increase accordingly. However, there is an upper limit to how complex the calibration model may be before the computational cost becomes prohibitive. A second limitation is related to the number of degrees of freedom of the model compared to the number of constraints that are available in the data. Regularization provides a feasible and computationally efficient solution to increase the sky and calibration model complexity without being restricted by the limited number of constraints.

\item Diffuse emission from the Galaxy is a major component that affects the short baseline interferometric data in addition to the discrete extragalactic sources. The construction of complete and accurate models for the diffuse sky is more difficult than for the discrete extragalactic sky. This is mainly due to the following reason: in order to model the diffuse structure accurately, we need to subtract the contamination from the discrete extragalactic sky, but this subtraction itself may create a form of suppression of the weak, diffuse structure. 
We also note that even with a complete model of the diffuse sky, calibration is not straight forward as most calibration methods assume discrete (groups of) components in the sky, but this has been overcome in recent work.
The exclusion of short baselines during calibration can also be used to our advantage. We can build a statistical jackknife test  (cross validation) by comparing the results with and without exclusion. This test can be done after processing all data, and without the need to have a separate telescope for cross validation.
Considering that LOFAR has longer baselines as well (even at VLBI scale), it is also necessary to exclude the longest baselines that are more affected by the ionosphere. These baselines are needed, however, to create high resolution sky models for bright sources.
The exclusion of short baselines has more impact than the exclusion of very long baselines in terms of effort and return in quality.

\item Regularization is essential in many data processing operations, including calibration. We can consider regularization (and maximum likelihood estimation) as a maximum, a-posteriori estimation with the prior defined by the regularizer. Hard regularization (with exact constraints) are always preferred to soft constraints that only act as a penalty. Most plausible priors enhance smoothness as physical processes including the sky and the instrument show such smooth behavior. However, this is not a free license to use smoothing indiscriminately. For example, separate smoothing (other than using as a regularizer) during or post-calibration may seem to improve the result but might not always work perfectly. In contrast, we can add regularizers that are not smooth, provided they describe the physical behavior of the systematics \citep{DCAL}.

\item Significant care needs to be given to stability and statistical efficiency of the algorithms in order to process thousand of hours of observations. Standard and well tested algorithms that have well defined convergence criteria should be adopted. It is also important to acknowledge that constant improvement/refinement of pipelines is necessary: for example, RFI mitigation methods that were deemed adequate a decade ago are no longer good enough for current interference mitigation, for example due to satellites  \citep{Cees2023}. 

\item The growth of complicated data processing algorithms accompany an equally complicated set of hyper-parameters to choose from. These hyper-parameters include various regularization factors and various basis functions (kernels). The fine-tuning of such hyper-parameters is a prerequisite to get the optimal performance from such complicated data processing algorithms. We can readily adopt modern machine learning techniques (like reinforcement learning or semi-supervised learning) for this purpose \citep{YA2021MNRAS,Mertens2024}
\end{itemize}


\subsection{Lessons learned from the HERA}
\label{sec:hera_lessons}

The HERA concept emphasizes collecting area at large angular scales at the expense of imaging performance.
The design is a regular array of large, close-packed dishes with many redundant measurements of the same visibilities to increase the power spectrum sensitivity \citep{Dillon2017}.
HERA Phase I featured a dish design combined with a caged dipole feed \citep{DeBoer2017}, and took data for a few years while the array was being built, culminating in improved upper limits on the 21\,cm power spectrum at $z \sim 8$ and $z \sim 10.4$ \citep{HERA2022a, HERA2023}.
For Phase II, the front-end system was upgraded to include a wideband Vivaldi feed 
that extends the instantaneous frequency coverage to $50-250$~MHz, and a new fiber analog system to remove coaxial cable reflections \citep{Berkhout2024}.
This system began science-grade observations in 2022, while being continuously rolled out.
Phase II is now mostly commissioned and data analysis has been actively underway since the first science observations.

The HERA design allows the experiment
to build up sensitivity by co-adding redundant baselines calibrated using the repeated measurements and time--filtering analysis approaches rather than imaging.
The redundant configuration has allowed the inspection of data repeatability at a fine-grained level which has been used to find subtle systematics that have been identified as areas for improvement in future designs.
Power spectra are formed without attempting to grid visibilities as in traditional imaging. This choice requires less stringent calibration requirements, but sacrifices some sensitivity compared to imaging power spectra.

Built exclusively for 21 cm cosmology, HERA elements such as the dish shape, feed matching and cable types were optimized for minimizing chromaticity \citep{Fagnoni21}.
The intrinsic spectral smoothness in an individual HERA element \citep{Neben2016, EwallWice2016} has been sufficient to achieve a demonstrated $>$4 orders of magnitude in foreground suppression, demonstrating the efficacy of the calibration and RFI excision. 
The biggest challenge facing HERA now is the additional chromaticity in the instrumental response from the electromagnetic interaction of elements \citep{Kern2019, Kern2020a, HERA2022a, HERA2023, HERAcoupling, Rath2025}.  



\subsection{Lessons learned from NenuFAR}
\label{sec:nenufar_lessons}

The NenuFAR Cosmic Dawn Key Science Program is designed to explore the redshifted 21-cm signal from neutral Hydrogen during the Cosmic Dawn ($15 \lesssim z \lesssim 30$, $30-85$~MHz). The programme targets large-scale 21~cm brightness temperature fluctuations predicted by models of the first stars and galaxies, and in particular provides a direct interferometric test of the unexpectedly deep absorption feature reported by EDGES at $z \sim 17$. NenuFAR is expected to reach the amplitude of the strongest Cosmic Dawn signals, those consistent with the EDGES absorption feature, after about 100~h of integration, and to probe standard scenarios after about 1000~h.

NenuFAR, located at the Nançay Radio Observatory (France) on the site of the French LOFAR station, operates as a stand-alone low-frequency interferometer and will ultimately be integrated into the LOFAR VLBI network as a super-station. Currently, the array comprises 80 Mini-Arrays (MAs) within a 400~m-diameter dense core, each containing 19 dual-polarised antennas laid out on a regular grid to simplify analogue beam-forming and delay-line implementation. This regular intra-MA geometry, while practical, generates strong grating lobes across the Cosmic Dawn frequency range. A small number of remote MAs, distributed up to a few kilometres from the core, provides intermediate baselines and improved imaging capability. The coaxial cables linking the MAs to the central container vary in length between 20 and 150~m, making it crucial to account for cable reflections through bandpass calibration to avoid artefacts at high delay (up to $3~\mu\mathrm{s}$), which can limit 21~cm cosmology analyses.

The observing strategy focuses on deep-field integrations. Initially, the North Celestial Pole (NCP) field was targeted, but a carefully selected high-elevation field (NT04) was later adopted to minimise foreground contamination. The data are processed through a dedicated pipeline combining \texttt{nenuprep} (pre-processing), \texttt{nenucal} (initial calibration), \texttt{NenuFlow} (direction-dependent calibration and source subtraction), and \texttt{pspipe} for power-spectrum estimation. Foreground separation is performed using ML-GPR, adapted from LOFAR-EoR processing. A more detailed description of the processing pipeline can be found in \cite{munshi2024first} and \cite{Munshi_improved_2025}.

The first years of Cosmic Dawn observations with NenuFAR have provided valuable insight into the challenges faced by compact low-frequency arrays operating in the $30-85$~MHz range. Three key lessons emerged: (1) impedance mismatches in the analogue chain generate cable reflections that need to be accurately modelled and mitigated in delay space; (2) the local RFI environment, though relatively stable, remains dynamic and dominated by intermittent ground-based emitters that can be identified through near-field imaging; and (3) the success of deep integrations depends critically on the selection of a clean, high-latitude field with minimal bright Galactic and extragalactic sources and high elevation. These lessons have direct relevance for the design and calibration strategy of SKA-low during its Cosmic Dawn (and EoR) campaigns.

\subsubsection{Cable reflections}

Cable reflections, caused by impedance mismatches at interfaces within the analogue signal chain (e.g., antennas, connectors, amplifiers), are not an uncommon issue in interferometers.
When a small fraction of the signal is reflected back and forth along a cable, it interferes with the direct signal, producing a standing wave whose period in frequency depends on the cable electrical length. In NenuFAR, each MA is connected to the central container through a cable of different length, so the associated reflection patterns appear at distinct delays. These quasi-periodic ripples imprint chromatic structures in the instrumental gains, coupling smooth foreground emission into high-delay modes that overlap with the 21~cm window. The different cable lengths induce ripples with delays ranging from approximately 0.3~$\mu$s to 3~$\mu$s.

In the NenuFAR Cosmic Dawn programme, these reflections are characterised and corrected using dedicated calibrator observations. A 30-minute observation of a bright source such as Cas~A or Cyg~A is taken before or after each target track. A direction-independent bandpass calibration is performed on this calibrator with a frequency resolution of 15.3~kHz, and the resulting gain solutions are applied to the target data. A subsequent calibration with a spectral-smoothness constraint sets the absolute flux scale. Delay-spectrum diagnostics confirm that this procedure effectively suppresses reflection peaks, minimising their impact on 21~cm power-spectrum measurements
\citep{munshi2024first}.

\subsubsection{Radio Frequency Interference}

Local RFI sources near NenuFAR were identified as a major contributor to the excess variance above the instrumental thermal noise in early analyses. The location and spectral characteristics of these emitters can be determined using near-field imaging techniques, which exploit the spherical-wavefront nature of local RFI to reconstruct three-dimensional source distributions \citep{munshi2025near}. Using this approach, one source was traced to air-conditioning units within the NenuFAR core with inadequate electromagnetic shielding, which has since been corrected. Such algorithms can be incorporated into real-time monitoring workflows at observatories to enable early identification and mitigation of local RFI. Improved modelling of these emitters could eventually allow subtraction of their contribution from the visibilities.

Terrestrial RFI sources add coherently at the celestial poles during rotation synthesis, since the poles remain stationary with respect to the array. As a result, RFI produces an artificial source at the pole in far-field images, whose PSF sidelobes contaminate the target field. Initial analyses of the NCP data were therefore strongly affected by RFI, as its sidelobe power spread across the $uv$ plane. In contrast, when the target field lies away from the poles, sidelobes can be efficiently filtered out.

\subsubsection{The importance of a carefully selected deep field}

Selecting an appropriate deep field is critical for maximising sensitivity and minimising systematics in 21~cm experiments. The initial NenuFAR Cosmic Dawn observations targeted the NCP, which offered continuous visibility and a stable primary beam~\citep{munshi2024first}. However, its fixed position also made it more vulnerable to stationary RFI sources and sidelobe leakage from bright A-team sources such as Cas~A and Cyg~A, which periodically transit through the beam pattern. These effects limited calibration stability and foreground separation. To overcome them, a systematic field-selection campaign was undertaken, 
targeting five candidate regions. The selection criteria included: (i) distance from the brightest sources like Cas~A and Cyg~A;
(ii) low diffuse Galactic emission; (iii) presence of a suitable in-field calibrator; and (iv) high average elevation 
to maximise sensitivity and suppress chromatic coupling of foregrounds within the delay ``wedge'' region~\citep{Munshi_beyond_2025}. Several fields were observed as test fields, with the final choice being labeled as NT04 field \citep[e.g.,][]{Munshi_beyond_2025}. This field was found to have lower RFI and sidelobe contamination compared to the NCP field, yielding a factor of about fifty improvement in power spectrum sensitivity~\citep{Munshi_improved_2025}.
%
%

\subsection{Lessons learned from MWA}
\label{sec:mwa_lessons}

The 
MWA is a low-frequency radio interferometer located at \textit {Inyarrimanha Ilgari Bundara}, the Commonwealth Scientific and Industrial Research Organisation (CSIRO) Murchison Radio-astronomy Observatory, on Wajarri Yamaji country, approximately 300 km from the nearest town \citep{Tingay2013}. This is the same remote, radio-quiet site where the SKA-low is being constructed.

The MWA has evolved through several development phases, each aimed at enhancing its scientific capabilities. In its initial phase (2013), the array comprised 128 tiles, each including 16 dipoles arranged in a $5 \times 5$~m regular grid, arranged in a pseudo-random configuration, optimized for high-fidelity imaging across a broad range of baseline lengths. During the second upgrade, the array was expanded to 256 tiles, with 72 of them arranged into two hexagonal sub-arrays to improve sensitivity to large-scale 21 cm cosmological signals by approximately a factor of 3.5, while the remaining 56 tiles were deployed on longer baselines to increase the $uv$ coverage \citep{Wayth2018}. However, the correlator at that time could process signals from only 128 antennas simultaneously, limiting the full use of the upgraded infrastructure. 
A new correaltor was deployed in the third upgrade, and the signal from 245 tiles can now be cross correlated simultaneously.

The MWA operates in both drift-scan and tracking modes. In drift-scan mode, the tiles observe the sky continuously as it drifts overhead with Earth’s rotation, similar to the observing strategy employed by HERA. In tracking mode, however, the tiles are digitally pointed at a specific sky position for a set duration before switching to the next pointing. These pointings are separated by approximately $6.8^{\circ}$, chosen to maintain a consistent primary beam response that aligns with the fixed analogue delay settings \citep{Tingay2013}.

\subsubsection{Lessons learned from tracking observations}  

For EoR observations, the MWA correlator performs cross-correlations with a frequency resolution of 10 kHz and a temporal cadence of 2 s. During the pre-processing stage, the data are typically averaged to 40 kHz in frequency and 8 s in time to reduce data volume while retaining sufficient spectral and temporal resolution for EoR analysis. Since the publication of the first MWA power spectrum results, significant efforts have been devoted to refining the calibration strategies, improving the foreground modelling, developing robust systematics mitigation techniques, and enhancing ionospheric corrections \citep{Jacobs2016, Beardsley2016, trott2016,Li2018, Trott2020, Rahimi2021, Kolopanis2023, Wilensky2023, Barry2024, Nunhokee2024, Nunhokee2025, Trott2025}. The latest results are consistent with a scenario where cold reionization is disfavoured \citep{HERA2023}
and provide the first evidence of IGM heating at $z = 6.5$–$7.0$ \citep{Nunhokee2025}. In this subssection, we summarise the key lessons learned from over a decade of operations in tracking mode:
 
\begin{itemize}
    \item The EoR observing programme targets three primary low-foreground fields: EoR0 (RA~=~$0^{\rm h}$, DEC~=~-27$^{\circ}$), EoR1 (RA~=~$4^{\rm h}$, DEC~=~-27$^{\circ}$) and EoR2 (RA~=~$10.3^{\rm h}$, DEC~=~-10$^{\circ}$). Among these, EoR0 has been identified as the optimal field for deep EoR observations due to its relatively low diffuse foreground emission and minimal contamination from bright extragalactic sources \citep{Trott2020}. However, it remains partially affected by Galactic plane emission during pre-zenith observations. In contrast, EoR1 and EoR2 suffer from contamination by bright radio sources, notably Fornax A in EoR1 and Hydra A in EoR2, which complicate calibration and foreground subtraction.
    
    \item Sources located within the main lobe of the primary beam, as well as those in its sidelobes, contribute significant power to the wedge region and even beyond the horizon in the cylindrical power spectra (Thyagarajan et al. 2015a,b; Trott et al. 2020; Pober et al. 2016). Given the MWA wide field of view ($\sim 26^{\circ}$ at $\lambda/2$), avoiding contamination from bright compact and extended sources remains challenging. \citet{Lynch2021} developed a high-fidelity foreground model for the EoR0 field, accurately capturing sources within both the main lobe and sidelobes of the primary beam across $100–230$~MHz, improving the foregroud separation.
    Additional efforts have focused on constructing shapelet-based models for extended sources such as Fornax~A \citep{Line2020}.
    
    \item The MWA employs a direction-dependent calibration scheme that uses least-squares algorithms to solve for the complex antenna gains \citep{Barry2019, Jordan2025}. These gains are derived independently for each frequency and time interval, across all four polarizations. The MWA correlator first channelizes the signal into 24 coarse channels, each 1.28~MHz wide, before further channelization to a finer resolution of 10~kHz. Within each coarse channel, foreground sources are modeled including the Full Embedded Element primary beam model \citep{Sokolowski2017}. While the direction-independent calibration results are promising, further improvements to the sky model or the application of self-calibration techniques could potentially enhance the calibration performance \citep{Jaiden2021}.
    
    \item Efforts to characterise and understand the instrument have been largely successful in identifying and mitigating instrumental systematics. 
    RFI remains a non negligible issue, and 
    a significant fraction of data discarded during processing is attributed to its contamination \citep{Nunhokee2024, Nunhokee2025}. Traditional RFI mitigation algorithms, such as AOFlagger and SSINS, tend to underperform when the interference is weak, buried under the noise, or not strongly amplitude-dependent \citep{Offringa2015, Wilensky2020}.
    
    \item Ionospheric corruptions are mitigated through a metric 
    that quantifies the quality of an observation in relation to ionospheric activity. In \citet{Trott2020}, an upper threshold for this metric was defined such that observations exceeding the optimal limit are classified as highly active and, consequently, excluded from further analysis. This threshold has since been incorporated into the data processing pipeline \citep{Nunhokee2025}.

    \item The MWA employs a hybrid foreground removal approach in which extragalactic sources are subtracted in the visibility domain, and only modes free from contamination are retained for constructing the final power spectrum. Direct subtraction of the foreground sky model alone proved insufficient to achieve satisfactory subtraction metrics. Consequently, phase offset corrections, due to ionospheric turbulence, were applied to a subset of bright sources to enhance the efficiency and accuracy of the subtraction process \citep{Nunhokee2024, Jordan2025}.

    \item Traditionally, two primary approaches have been used to perform the power spectrum analysis. The first is the Cosmological HI Power Spectrum Estimator (CHIPS), which grids and averages the visibilities directly in the visibility domain. 
    The second approach, Fast Holographic Deconvolution (FHD/$\epsilon$ppsilon), involves transforming the data into the image domain for gridding and subsequent power spectrum estimation \citep{Barry2019}. While both algorithms have their respective limitations, their results have been found to be consistent on large angular scales \citep{Nunhokee2025}.

    \item One of the key limitations of the MWA correlator was the use of a polyphase filter bank that required flagging of the edge and center channels of each coarse frequency band. This process introduced harmonic structures in the Fourier space, contributing excess power within the wedge region. This limitation has been addressed in the third phase of the MWA upgrade, resulting in a significant reduction of power at these coarse channel harmonics in the power spectra.
\end{itemize}

The MWA data processing pipeline is designed to efficiently manage large data volumes and is scalable to meet the demands of future 21~cm experiments with SKA-low.


\subsubsection{Lessons learned from drift scan observations} 
\label{sec:mwa_drift}

In this section, we discuss drift scan observations, where the telescope pointing direction remains fixed relative to the Earth and, therefore, changes continuously in the celestial reference frame due to the Earth rotation. The fact that the telescope primary beam remains constant in drift scan observations facilitates a
foreground characterization across the long observations required to detect the cosmological 21~cm signal. 

Novel techniques were developed to analyze MWA drift scan observations and tackle the magnitude of the foreground contamination
\citep{Ali2008, Bernardi2009, Ghosh2012, Paciga2013}. 
\citet{Chatterjee2022} introduced the  Tracking Tapered Gridded Estimator (TTGE) for estimating the 21~cm power spectrum from drift scan observations. The TTGE is a generalisation of the existing Tapered Gridded Estimator  \citep[TGE,][]{Choudhuri2014, Choudhuri2016b}, a visibility-based, 21-cm power spectrum estimator that suppresses the sidelobe responses of the telescope to mitigate the effects of extra-galactic foregrounds \citep{Ghosh2011a, Ghosh2011b}. Additionally, the TGE is computationally efficient as it deals with gridded visibilities. It is also an unbiased estimator as it either internally estimates the noise bias from the self-correlation of the visibilities \citep{Pal22} or uses a cross-correlation between two orthogonal polarizations \citep{Elahi2023} to yield an unbiased power spectrum estimate.

Missing frequency channels, flagged to remove RFI, also pose a serious problem for visibility-based power spectrum estimation as the Fourier transform from frequency to delay space \citep[e.g.,][]{Morales2004, Parsons2009} introduces power spectrum artefacts. To overcome the issue in the analysis of drift scan observations, the multi-frequency angular power spectrum $C_{\ell}(\nu_a,\nu_b)$ (MAPS; \citealt{Datta2007,mondal2018}) is first estimated, then 
$C_{\ell}(\nu_a,\nu_b) = C_{\ell}(\Delta \nu)$, where $\Delta \nu=\mid \nu_a-\nu_b \mid$, is computed. In the TTGE,  visibilities are first correlated to estimate $C_{\ell}(\dnu)$, which is, in turn, Fourier transformed along the frequency direction 
to estimate $P(k_\perp, \kpar)$, the cylindrical power spectrum \citep{2019MNRAS.483.5694B}.


MWA observations have a periodic pattern of flagged channels which introduce
horizontal streaks in the power spectrum 
\citep{Paul2016, Li2019, Trott2020, Patwa2021}. 
%
This problem was addressed by 
\citet{Elahi2025} 
by introducing smooth component filtering \citep[SCF,][]{Elahi2025}, that removes the slowly varying spectral component of the measured gridded visibility data. The combination of the SFC and TTGE analysis yields to an improvement on earlier results, leading to a $(934.60)^2 \, {\rm mK^2}$ upper limit at $k=0.418 \,{\rm Mpc}^{-1}$. 
\subsubsection{Bispectrum from MWA drift scan observations} 
\label{sec:mwa_bs}

The bispectrum, a higher order statistic sensitive to non-Gaussianities, is expected to provide substantial insights into the EoR that are inaccessible by the power spectrum \citep{Bharadwaj2005, sumanm2020, kamran2021, Watkinson_2022, gill_eormulti}. However, estimating the bispectrum from radio-interferometric data is computationally challenging due to its high dimensionality. Recently, \citet{Gill_2025_est} have proposed an efficient estimator of the 21 cm bispectrum that operates on gridded visibilities and leverages the FFT-based acceleration. \cite{Gill_2025} have applied the estimator to the MWA drift scan observations at $z=8.2$, and the measurements of the three-dimensional bispectrum, finding a foreground wedge feature in the $k_{1\parallel}-k_{1\perp}$ plane, very similar to the cylindrical power spectrum.
The bispectrum also exhibits a clear EoR window, where all three triangle sides are outside the foreground wedge, and upper limits on the average mean brightness temperature fluctuations are derived at $(2.04\times10^{3})^3~{\rm mK}^3$ level \citep[see][for details]{Gill_2025}.


\subsection{Lessons learned from uGMRT observations}
\label{sec:gmrt_lessons}

Over the past decade, the upgraded GMRT \citep{2017CSci..113..707G} has served as a valuable precursor to the SKAO. Operating between 120 and 1400~MHz with wide-band receivers and a digital backend \citep{2017JAI.....641011R}, the uGMRT bridges the frequency gap between SKA-low and SKA-mid. The experience with uGMRT Band-2 ($120-250$\,MHz) provides key lessons in observation design, calibration, and foreground characterization for SKA-low.

\begin{itemize}
\item The uGMRT Band-2 ($120-250$\,MHz) is affected by radio frequency interference (RFI) from communication transmitters, power infrastructure, and satellites. Automated RFI detection and flagging are essential for preserving data quality. RFI are mitigated using a multi-layered strategy that combines hardware filtering, real-time digital excision, and post-correlation flagging \citep{2023JApA...44...37B}.

\item Observations of the ELAIS-N1 field with uGMRT Band-2 led to foreground characterization as well as constraints on the 21~cm signal.
Fields at high Galactic latitudes were chosen to minimize bright sources and diffuse foregrounds \citep{2017A&A...598A..78I,2020MNRAS.494.1936C}, improving calibration convergence and sensitivity to the faint 21~cm signal.

\item Ionospheric phase fluctuations pose a major challenge at low frequencies. They cause rapid changes in visibility phases and can decorrelate signals on long baselines within minutes \citep{2025JCAP...02..058P}. The SPAM pipeline \citep{2014ASInC..13..469I,2017A&A...598A..78I} performs direction-dependent calibration and ionospheric phase correction, which substantially improves continuum image fidelity \citep{2025arXiv251102375S}.

\item Wide-band observations ($120-250$\,MHz) enhance sensitivity to diffuse structures but require careful frequency-dependent calibration. Beam chromaticity and instrumental gain variations must be corrected to avoid frequency-dependent leakage of spectrally smooth foregrounds into the 21~cm signal \citep{2022MNRAS.512..186K, 2024JCAP...05..068G,2025JCAP...07..024G}. An accurate primary-beam model is therefore crucial for maintaining spectral smoothness.

\item Wide-field imaging at low frequencies is impacted by the frequency- and direction-dependence of the primary beam. Frequency-dependent primary beam correction and w-stacking  algorithms are important to keep flux densities and spectral indices consistent across the field \citep{2014MNRAS.444..606O}. \texttt{WSClean}\footnote{\url{https://gitlab.com/aroffringa/wsclean}} was the package routinely used for imaging.
Source catalogues were compared with LoTSS \citep{2021A&A...648A...2S} 
to check flux accuracy, positional offsets, and source counts. After beam and calibration corrections, uGMRT flux densities are consistent with published catalogues within typical catalog uncertainties \citep{2025arXiv251102375S}.

\item 
Short uGMRT baselines are expecially suited to study diffuse Galactic emission.
Galactic brightness fluctuations were characterised statistically in both image and visibility domains using the angular power spectrum 
\citep{2014MNRAS.445.4351C,2016MNRAS.463.4093C}. This dual-domain analysis confirms a spatial power-law behaviour of the Galactic emission across $120-500$\,MHz \citep{2025arXiv251102375S}. Power-spectrum analyses of diffuse emission \citep{2020MNRAS.494.1936C} show that residual calibration and beam-model errors introduce excess power on small angular scales, providing further need to mitigate these effects in 21~cm observations.

\item The approach followed in the analysis of the 21~cm uGMRT observations is hybrid, not too different from the LOFAR an MWA approaches. Compact sources are first subtracted from the visibility data, then the residual power spectrum is computed by masking out the wedge region \citep{2010ApJ...724..526D,2012ApJ...753...81P}, together with a principal component separation of residual foreground emission \citep{2012MNRAS.419.3491L,2012MNRAS.423.2518C}. This strategy has been found effective and has been tested on simulated SKA-low observations, effectively improving the recovery of the 21~cm signal \citep{2025JCAP...10..035T, 2025arXiv251025886B}. 
\end{itemize}

\subsubsection{On the use of uGMRT observations to calibrate other EoR arrays}
\label{sec:uGMRT_eor_cal}


Given its $120 - 1500$~MHz frequency coverage and the good angular resolution ($\sim$20 arcsec at 150~MHz), the uGMRT is a unique instrument to provide foreground characterization for 21~cm observations, including other instruments that have, for example more limited angular resolution.

A preparatory analysis\footnote{Code here: \url{https://github.com/aecosmo/RADIOcat}} has demonstrated that the calibration of HERA observations can actually be improved with a high resolution sky model derived from uGMRT observations.
A catalogue of sources down to a 20~mJy flux density of the GLEAM~02H (J0200–3053) field, used to calibrate HERA observations \citep{Kern2020b, HERA2022a}, is publicly available \citep{2025MNRAS.544..321E}.


\section{Spectral response of the SKA-low log periodic antenna and their impact on CD/EoR observations}
\label{sec:skala}

The SKAO log-periodic antenna (SKALA) has been carefully designed to deliver broadband, spectrally smooth performance over the full 50--350~MHz band required by CD and EoR observations. Nevertheless, several subtle instrumental effects can introduce chromatic structures -- often referred to as ``spectral artefacts'' -- that may compromise the exquisite spectral smoothness needed to isolate the faint 21~cm signal from bright astrophysical foregrounds. These artefacts arise from a combination of electromagnetic, mechanical, and signal-chain
phenomena, which we outline below.
%
While there is a section in this book addressing mutual coupling effects primarily associated with the array layout and inter–element interactions, the focus here shifts to spectral artifacts that originate within the antenna system itself.
These intrinsic, element–level effects
can also introduce chromaticity that impacts CD/EoR observations with SKA--low.

The SKALA antenna programme began at the Cavendish Laboratory, University of Cambridge, in 2009 as part of the early aperture--array technology development for SKA--low. Initial prototypes (SKALA0 and SKALA1) demonstrated the feasibility of a dual--polarised, log--periodic dipole array covering the $50-350$~MHz band with high sensitivity and wide field--of--view \citep{deleraacedo2015skala}. Subsequent generations (SKALA2 and SKALA3) introduced refinements to the dipole geometry, balun, and low--noise amplifier to improve impedance matching, spectral smoothness, and mechanical robustness \citep{SKALA3}. The design culminated in SKALA4 \citep{SKALA4_ref1, SKALA4_ref2, LFAA_station}, reviewed and accepted as the baseline element for SKA--low, followed by the SKALA4.1 \citep{SKALA41} and the current SKALA4.2 versions, which incorporate optimisations for manufacturability and mechanical durability. These later iterations led to the production of an industry--ready design by \emph{SIRIO Antenne S.r.l.} and \emph{INAF}, now responsible for large--scale fabrication of SKA--low antenna elements.

From an electromagnetic design perspective, the principal modification between SKALA4 and the later SKALA4.1 and SKALA4.2 versions is the transition from a pseudo--differential low--noise amplifier (LNA) to a single--ended LNA, together with the direct electrical grounding of the antenna boom to the station ground mesh. The adoption of a single--ended LNA brings the input impedance closer to 50~$\Omega$, but requires the inherently high--impedance, log--periodic structure to be matched to a lower load impedance. This adjustment modifies the current distribution along the boom and dipoles, potentially introducing small impedance discontinuities that lead to residual standing currents within the antenna arms. These internal reflections can generate weak standing--wave patterns that manifest as subtle spectral features in the antenna response.  

The second modification, creating a galvanic connection between the antenna boom and the ground mesh, is hypothesised to alter the current return path in a way that supports a monopole-like or parasitic mode below 100 MHz. This mode, which is only weakly coupled at the feed port of an isolated antenna, may become strongly excited in the array environment through lateral illumination and mutual coupling. The resulting additional current path could then produce a narrowband radiation feature, observed as a shallow low-frequency notch in the beam pattern (see Fig.~\ref{fig:notche_effect_on_BIR}). While this effect can be potentially mitigated through careful calibration or by restricting analysis to frequency ranges unaffected by the notch, it highlights the sensitivity of wideband aperture arrays to small grounding and impedance variations. If not compensated for, such narrowband artefacts could impact CD/EoR observations by introducing beam chromaticity.

\subsection{Impedance matching between the antenna and the LNA}

Achieving smooth impedance matching across the 7:1 frequency range of SKA--low ($50 - 350$~MHz) remains one of the most demanding aspects of wideband array operation. Small impedance discontinuities between the dipoles, balun, and low--noise amplifier introduce partial reflections that establish standing waves along the feed network. These reflections imprint highly coherent, quasi--sinusoidal ripples on the antenna voltage gain, producing periodic structures in both the passband and the measured visibilities. Such chromatic ripples translate directly into delay--space power, contaminating the power spectrum region that is expected to be foreground free.

\citet{SKALA3} and \citet{SpectralII} showed that even low--level mismatches can generate residual spectral structure well above the thermal noise floor if unmodelled, leading to foreground leakage across the EoR window. Although design optimisations (e.g.\ from SKALA2 to SKALA3) have improved the intrinsic passband smoothness, complete suppression of reflection--induced artefacts is neither realistic nor required for successful science operations. Instead, the emphasis now shifts toward calibration and mitigation strategies that quantitatively model these effects and incorporate them into the SKA--low data processing pipeline.  
In particular, the use of low--order polynomial bandpass models combined with direction--dependent calibration has been demonstrated to reduce reflection residuals to levels consistent with the tolerances derived for EoR detection---namely fractional amplitude residuals below $10^{-3}$ over a few MHz and phase variations under 0.04° per fine channel. Future operations will therefore take advantage of frequent, high signal--to--noise bandpass calibration, together with joint modelling of instrumental and sky spectral structure, to ensure that impedance--related chromaticity remains well understood and can be accounted in the power spectrum analysis. 

\subsection{Finite ground plane and soil effects}

Recent full–wave analyses have shown that the ground and underlying soil not only affect the electromagnetic response of the array \citep{SKA_soil_fastsim} but also contribute significantly to its system noise. Thermal emission from the lossy, multilayered soil can couple into the antenna ports, introducing direction–dependent noise correlations between elements and modulating the system temperature by several kelvin depending on soil permittivity, moisture, and temperature gradients \citep{SKA_soil}.  
These effects vary with antenna position and ground–plane geometry and are enhanced at low elevations where the interference between direct and ground–reflected fields is strongest.  
Accurate electromagnetic and thermal modelling of the layered medium is therefore important not only to predict spectral artefacts but also to quantify the soil–induced noise coupling that directly
impacts the sensitivity of SKA--low.

\subsection{Narrowband beam artefacts}

\begin{figure}[htb!]
\centering
\includegraphics[width=0.75\linewidth]{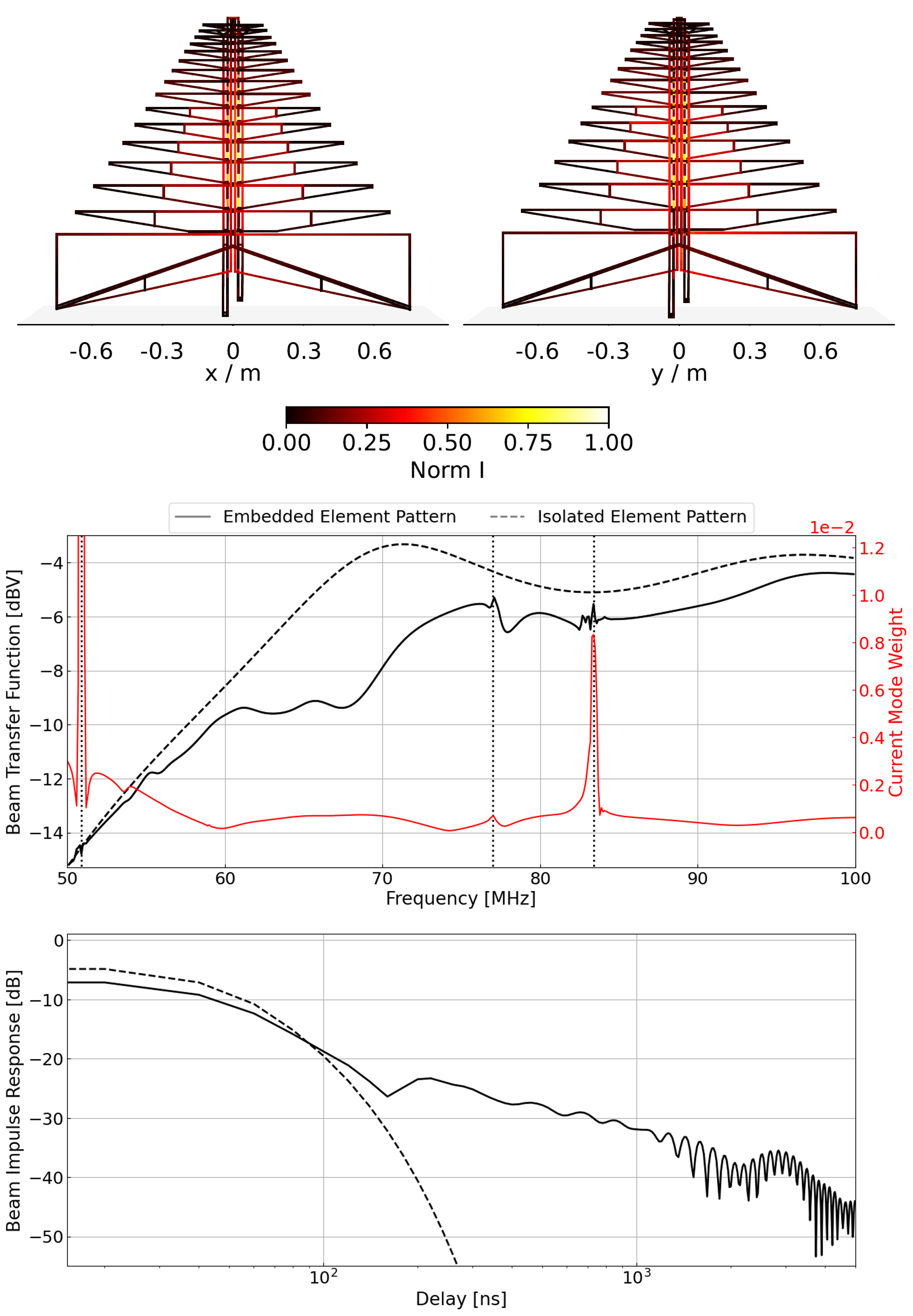}
\caption{Top: the SKALA4 antenna geometry illustrating the $7^{\textrm{th}}$ current mode, corresponding to a boom standing wave that is thought to be induced by element cross-coupling. Middle: the transfer functions of the isolated element pattern (IEP, dashed) and embedded element pattern (EEP, solid) at zenith are shown in black for the central antenna in a perturbed Vogel station layout. The red curve shows the weighting of the illustrated current mode for the EEP, where notches and glitches appear in the frequency band as this mode is excited. Bottom: the beam impulse response in delay space for the IEP and EEP, demonstrating the impact of the mode-induced notches at late delays \citep{cumner_cma,ohara_modelling}.}
\label{fig:notche_effect_on_BIR}
\end{figure}

The physical dimensions of the SKALA -- a large, multi--dipole, log--periodic structure -- can lead to narrowband beam resonances arising from the constructive and destructive interference of multiple current modes within the antenna arms. These features appear as small oscillatory variations in the element and station beams, introducing localised spectral structure that can bias foreground--subtraction algorithms. While these artefacts are weaker than those produced by mutual coupling between elements, they are non--negligible for wideband analyses that rely on high beam smoothness.  

Electromagnetically, these resonances originate from impedance mismatches between the array of dipoles, the transmission line along the boom, and the LNA located at the feed point. Such mismatches cause partial reflections and residual currents to appear in dipoles that would otherwise be electrically inactive at a given frequency, effectively exciting unwanted radiation modes within the structure. This behaviour is particularly difficult to eliminate when matching a low--impedance LNA to the intrinsically high--impedance ultra--wideband 
log--periodic antenna, making some degree of spectral ripple unavoidable. The problem is further exacerbated by electromagnetic coupling between the two orthogonal polarisation arms, which introduces additional mode mixing and spectral fine structure in both co--polar and cross--polar responses.  

In addition, pronounced resonances arising from standing waves along the antenna boom can give rise to monopole-like current distributions, particularly when the boom is galvanically connected to the station ground mesh, as in SKALA4.2. These effects can be further enhanced by lateral illumination and mutual coupling within the array environment, leading to the excitation of low-frequency parasitic modes (see Fig.~\ref{fig:notche_effect_on_BIR}).

Careful electromagnetic characterisation and cross--validation with measurement campaigns are underway to quantify these effects across the SKA--low band. As illustrated in \citet[][Figs.~5~and~7]{SKALA41}, such narrowband anomalies are clearly visible in the beam response across the frequency band, underscoring the need for precise modelling and calibration.

\subsection{Cable reflections and signal-chain chromaticity}

Reflections within the analogue signal path represent another important source of spectral artefacts in the SKA--low system. Even small impedance mismatches along transmission lines, connectors, or at LNA interfaces produce partial reflections that form standing waves within the coaxial cables. These reflections generate quasi--periodic ripples in the bandpass response, with frequencies set by the electrical length of the cables -- typically of the order of a few~MHz. Such ripples imprint coherent chromatic structure in the instrument gain, translating into localized power excesses at discrete delays that contaminate otherwise clean power spectrum regions.

Recent end--to--end simulations have shown that even sub--percent deviations in cable length or impedance can cause reflection amplitudes large enough to bias power spectrum measurements \citep{SKA_cables}. These effects are particularly insidious because they couple smooth foreground emission into high delay modes,
where the cosmological signal is expected to dominate. Mitigating them requires rigorous impedance control across all analogue components, careful cable equalisation and routing, and calibration strategies capable of modelling the frequency--dependent reflection terms to high precision. Without such mitigation, residual reflection signatures can persist across the full observing band, posing a significant challenge for high--fidelity CD/EoR experiments.

%

\section{Investigation of mutual coupling effects in the SKA-low layout}
\label{sec:mutual}
\label{sec:mutual}




SKA--low stations are composed of 256  closely-spaced log–periodic antennas, and the electromagnetic interaction between them -- \emph{mutual coupling} -- may become a significant source of spectral structure. Mutual coupling modifies the embedded element patterns (EEPs), introduces frequency–dependent structure in the mainlobe and sidelobes, and causes polarisation leakage. If not modelled accurately, these effects can contaminate measurements of the cosmological signal. In this section we will present an analysis of the impact of mutual coupling on 21~cm observations.

The 256 dual--polarised SKALA elements are arranged within a 40~m diameter phased--array station (see Section~\ref{sec:skala}).
The configuration aims for a high filling factor to achieve the required field of view and brightness sensitivity, which in turn imposes a tight inter--element spacing. Regular arrays are known to suffer from scan--blindness and other mutual--coupling--induced anomalies across wide frequency bands. To mitigate these effects, an initially proposed pseudo--random layout was adopted to minimise mutual coupling, later replaced by a deterministic Vogel (``sunflower'') layout that allowed slightly larger inter--element spacing. However, the regularity of the Vogel geometry produced a deep null in both element and station beams around 125~MHz, a signature of coherent coupling between elements. An optimisation campaign was performed to mitigate this anomaly and prevent similar effects across the observing band, effectively reverting to a \emph{perturbed Vogel} (PV) or pseudo--random configuration~\citep{AnsteySKA}. Randomised layouts distribute coupling over many spatial modes and frequencies, transforming strong discrete resonances into a broadband ``noise floor'' similar to sidelobe behaviour. Nevertheless, the residual chromaticity introduced by this distributed coupling still produces foreground leakage into the EoR window, compromising foreground--avoidance strategies if not accurately modelled. Additionally, mutual coupling between adjacent stations -- particularly within the dense SKA--low core -- cannot be neglected \citep{Quentin1}, as inter--station coupling is expected to add further chromatic structure that must either be calibrated or properly understood when applying avoidance approaches. Similar effects have been observed in compact arrays such as HERA, where coupling signatures extend across the entire EoR window~\citep{HERAcoupling,Rath2025}.

\subsection{Full EM simulations and mutual coupling}

Mutual coupling in closely packed aperture arrays causes each element to radiate a different EEP depending on its position in the array and its interaction with neighbours~\citep{MutualCoupling}. Simplified analytic models are unable to capture the resulting angular and spectral structure. To address this, a comprehensive simulation framework based on a fast full–wave EM solver (\emph{HARP},~\citealt{David, Quentin0, Bui_Van_2018}, later \emph{FAST},~\citealt{Quentin1, Quentin2}) and the GPU–accelerated radio telescope simulator \emph{OSKAR}~\citep{dulwich_2020_3758491} was developed for SKAO observations since 2009. These tools generate embedded element patterns for arbitrary array layouts and propagate them through to visibilities. The \emph{FAST} code is capable of performing full--wave electromagnetic simulations of SKA--low stations several orders of magnitude faster than conventional commercial solvers, which are typically optimised for generality rather than for the specific array configurations encountered in radio astronomy. \emph{FAST} is designed with flexibility and scalability as primary goals, allowing efficient modelling of complex, wide--band, dual--polarised aperture arrays such as those used in SKA--low. It has also been successfully benchmarked on simulations of up to fifty full SKA--low stations, demonstrating scalability to the entire SKA--low core and beyond.

\subsection{Impact of mutual coupling on 21~cm observations}


\begin{figure}[htb!]
\centering
\includegraphics[width=0.8\linewidth]{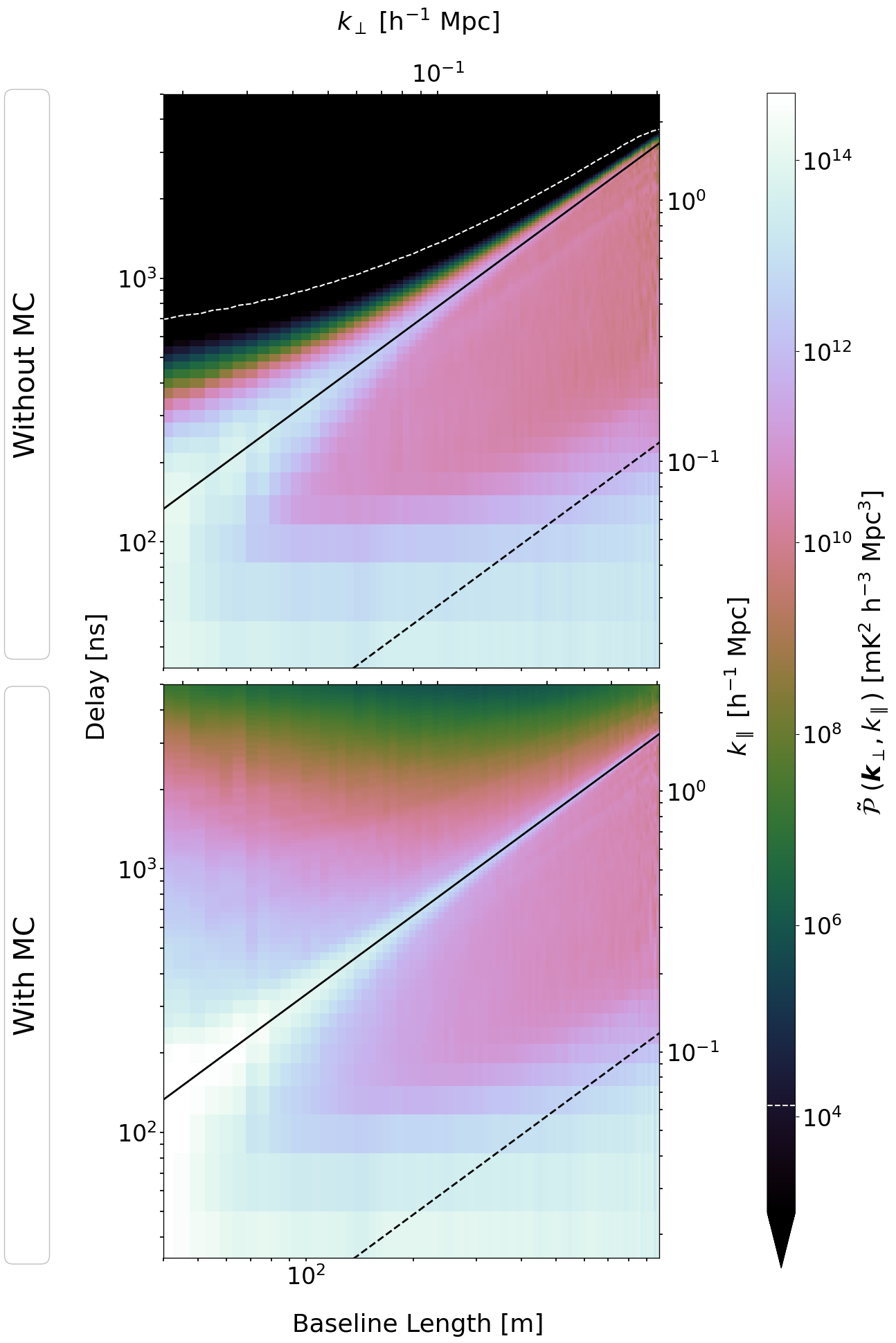}
\caption{Delay power spectrum for the SKA--low core with a random station layout over the 120--150~MHz band at 100~kHz resolution. The solid black line marks the horizon limit, the dashed black line indicates the beam limit, and the dashed white line shows the foreground spill-over. Mutual coupling broadens the delay impulse response, causing band-limited foregrounds —- typically confined within the horizon limit —- to contaminate the entire accessible detection window when compared to the isolated element pattern. Figure taken from \citet{ohara2025uncoupling}.}
\label{fig:eep_vs_iep}
\end{figure}

\citet{ohara2025uncoupling} simulated regular, sunflower and pseudo-random SKA--low station layouts in the $120 - 150$~MHz band with a 100~kHz resolution and found that mutual coupling significantly increases foreground leakage at high delays and obscures the 21~cm detection window for all layouts, as shown in Fig.~\ref{fig:eep_vs_iep}.
When the foreground contribution is subtracted using approximate beam models based on average or coarsely sampled embedded element patterns, residuals still remain at $\sim$1\,\% level, 
several orders of magnitude brighter than the expected EoR signal. Although these simulations should be interpreted as a \emph{worst--case scenario} and
more sophisticated calibration and modelling approaches can partially mitigate these effects -- for instance, through hybrid strategies combining direction-dependent calibration with
foreground avoidance 
-- they highlight the need to accurately investigate the effect of mutual coupling on beam patterns.
Even under favourable conditions, beam models may need to be known
to at least 4--5 significant digits in the far--field response and at high spectral resolution to ensure sufficiently smooth behaviour for robust foreground subtraction and to prevent residual leakage into the EoR window.

%
%

\section{Conclusions}
\label{sec:conclusions}

\begin{figure}
    \centering
    \includegraphics[width=\linewidth]{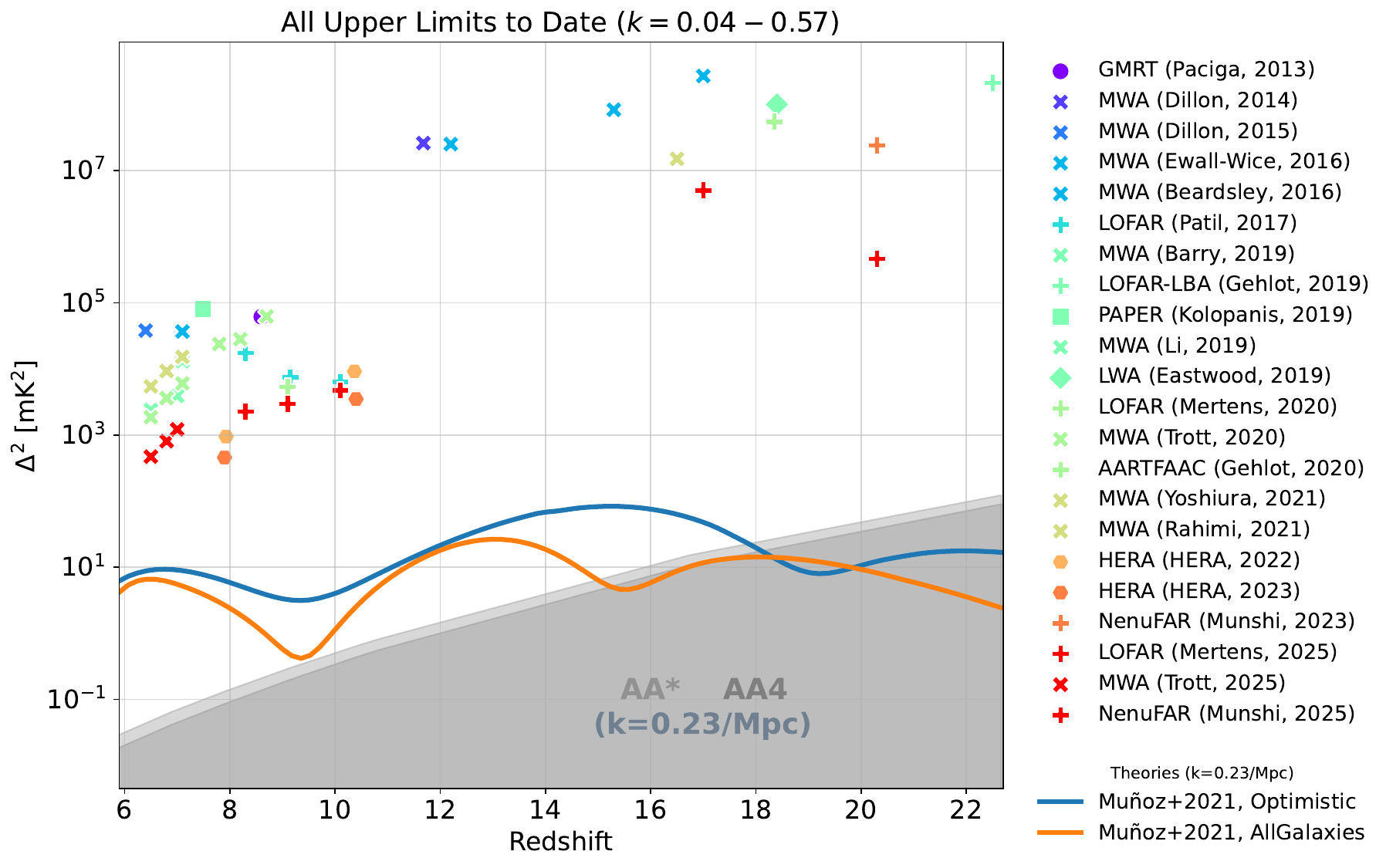}
    \caption{Current status of power spectrum upper limits from all 21\,cm experiments \citep{Paciga2013,Dillon14,Dillon15,Ewall-Wice16a,beardsley_first_2016,Patil2017,Barry2019b,Gehlot19,Kolopanis19,Li2019,Eastwood19,mertens2020,Trott2020,Gehlot20,Yoshiura21,Rahimi2021,HERA2022a,HERA2023,mertens2025,Trott2025,Munshi_improved_2025}. Different marker types indicate different experiments, while color indicates the year of publication (the legend is ordered by publication date). Also included are two theoretical predictions from \citet{Munoz2022}, and the expected thermal sensitivity of one year (1000 hours) of observations with SKA-LOW in both AA$^*$ and AA4 configurations, computed with \texttt{21cmSense} \citep{Pober2014,Murray2024}.}
    \label{fig:current-status}
\end{figure}

In this chapter we reviewed more than a decade of observations of the redshifted 21~cm line with SKAO pathfinders and precursors. We present a single-plot summary of how upper limits have been steadily decreasing over years, reflecting the sensitivity improvement of the observations, as well as the data analysis effort of the community (Figure~\ref{fig:current-status}). Based on this experience, which lessons can be drawn for future SKA-low 21~cm observations?
Although there are not definitive answers yet, a few points have emerged from the different instruments and analysis approaches:
\begin{itemize}
    \item end-to-end simulations and validation pipeline have become more and more a standard across the various experiments and we expect this to continue in the SKAO era;
    \item although the development of calibration and foreground separation methods and strategies has become mature and effective in dealing with foreground emission, even when distorted by a non-smooth instrumental response, its coupling between a poorly known instrumental response -- either in the form of primary beam far-sidelobes, cable reflections, electronic cross-talk, or other effects -- may be the limiting factor in current observations and will likely remain relevant in future SKA--low observations. The importance of mutual coupling effects has been increasingly recognized in the community: mutual coupling effects in closely packed antennas within an SKA--low station will require dedicated modeling effort and mitigation strategies in order to detect the 21~cm signal;
    \item despite the improvement in RFI mitigation techniques, RFI contamination is worsening over time and low-level RFI is a concerning systematic that will require attention in SKA--low observations;
    \item ionospheric distortions have been mitigated in current 21~cm observations and may not appear to be a showstopper to the detection of the 21~cm in future SKA-low observations;
    \item observations with pathfinders and precursors have highlighted the benefit of creating more accurate sky models using high angular resolution observations for both calibration and foreground separation purposes.
\end{itemize}

\section*{Author contributions}

Eloy de Lera Acedo led the write up of Section~\ref{sec:skala} and Section~\ref{sec:mutual}.
James Aguirre contributed to Section~\ref{sec:simulations_validation} and Section~\ref{sec:hera_lessons}.
Dominic Anstey contributed to Section~\ref{sec:skala} and Section~\ref{sec:mutual}.
Nichole Barry contributed to the body of text in Section~\ref{sec:simulations_validation} via introducing the concept, detailing specific citable examples, and drawing themes across all collaborations.
Gianni Bernardi led the chapter development, coordinated its overall writing and contributed to it.
Somnath Bharadwaj contributed to the writing of Section~\ref{sec:mwa_drift}.
Anthony Brown contributed to Section~\ref{sec:skala} and Section~\ref{sec:mutual}.
Jean Cavillot contributed to Section~\ref{sec:skala} and Section~\ref{sec:mutual}.
Suman Chatterjee contributed to the writing of Section~\ref{sec:mwa_drift}.
Samir Choudhuri contributed to the writing of Section~\ref{sec:mwa_drift}.
Tyler Cox contributed to Section~\ref{sec:simulations_validation}.
John Cumner contributed to Section~\ref{sec:skala} and Section~\ref{sec:mutual}.
Abhirup Datta contributed to Section~\ref{sec:gmrt_lessons}.
Fred Dulwich contributed to Section~\ref{sec:skala} and Section~\ref{sec:mutual}.
Khandakar Md Asif Elahi contributed to the writing of the Sections~\ref{sec:mwa_drift}~and~\ref{sec:uGMRT_eor_cal}.
Andrew Faulkner contributed to Section~\ref{sec:skala} and Section~\ref{sec:mutual}.
Sukhdeep Singh Gill contributed to the writing of Section~\ref{sec:mwa_bs}
Quentin Gueuning contributed to Section~\ref{sec:skala} and Section~\ref{sec:mutual}.
Daniel C. Jacobs contributed to Section~\ref{sec:simulations_validation}.
Nicholas Kern contributed to Section~\ref{sec:simulations_validation} and Section~\ref{sec:pathfinders_precursors}
Piyanat Kittiwisit contributed to Section~\ref{sec:simulations_validation}.
Yuchen Liu contributed to Section~\ref{sec:skala} and Section~\ref{sec:mutual}.
Zachary Martinot  contributed to Section~\ref{sec:simulations_validation}.
Ashish Mhaske contributed to Section~\ref{sec:skala} and Section~\ref{sec:mutual}.
Florent Mertens contributed to Section~\ref{sec:nenufar_lessons}.
Vincent McKay contributed to Section~\ref{sec:simulations_validation}.
Satyapan Munshi contributed to Section~\ref{sec_pipeline_validation} and Section~\ref{sec:nenufar_lessons}.
Steven Murray developed the structure, and wrote the text for much of the discussion in Section~\ref{sec:simulations_validation}.
Chuneeta D. Nunhokee  contributed to the writing of Section~\ref{sec:mwa_drift}.
Oscar Sage David O'Hara contributed to the writing of Section~\ref{sec:skala} and Section~\ref{sec:mutual}, and the preparation of the associated figures.
Samit K. Pal contributed to Section~\ref{sec:gmrt_lessons}.
Robert Pascua contributed to Section~\ref{sec:simulations_validation}.
Rashmi Sagar contributed to Section~\ref{sec:gmrt_lessons}.
Shouvik Sarkar contributed to the writing of Section~\ref{sec:mwa_drift}.
Shiv Sethi contributed to the writing of Section~\ref{sec:mwa_drift}.
Sarod Yatawatta contributed to Section~\ref{sec:lofar_lessons}.
Oskar Zetterstrom contributed to Section~\ref{sec:skala} and Section~\ref{sec:mutual}.




\bibliographystyle{abbrvnat}
\bibliography{chapter} 

@INPROCEEDINGS{2014ASInC..13..469I,
       author = {{Intema}, H.~T.},
        title = "{SPAM: A data reduction recipe for high-resolution,low-frequency radio-interferometric observations}",
    booktitle = {Astronomical Society of India Conference Series},
         year = 2014,
       series = {Astronomical Society of India Conference Series},
       volume = {13},
        month = jan,
        pages = {469},
archivePrefix = {arXiv},
       eprint = {1402.4889},
 primaryClass = {astro-ph.IM},
       adsurl = {https://ui.adsabs.harvard.edu/abs/2014ASInC..13..469I}
}

@ARTICLE{2017A&A...598A..78I,
       author = {{Intema}, H.~T. and {Jagannathan}, P. and {Mooley}, K.~P. and {Frail}, D.~A.},
        title = "{The GMRT 150 MHz all-sky radio survey. First alternative data release TGSS ADR1}",
      journal = {\aap},
         year = 2017,
        month = feb,
       volume = {598},
          eid = {A78},
        pages = {A78},
          doi = {10.1051/0004-6361/201628536},
archivePrefix = {arXiv},
       eprint = {1603.04368},
 primaryClass = {astro-ph.CO},
       adsurl = {https://ui.adsabs.harvard.edu/abs/2017A&A...598A..78I}
}

@ARTICLE{2022MNRAS.512..186K,
       author = {{Kumar}, Jais and {Dutta}, Prasun and {Choudhuri}, Samir and {Roy}, Nirupam},
        title = "{Calibration requirements for Epoch of Reionization 21-cm signal observations - II. Analytical estimation of the bias and variance with time-correlated residual gains}",
      journal = {\mnras},
         year = 2022,
        month = may,
       volume = {512},
       number = {1},
        pages = {186-198},
          doi = {10.1093/mnras/stac499},
archivePrefix = {arXiv},
       eprint = {2202.09589},
 primaryClass = {astro-ph.IM},
       adsurl = {https://ui.adsabs.harvard.edu/abs/2022MNRAS.512..186K}
}

@ARTICLE{2025JCAP...10..035T,
       author = {{Tripathi}, Anshuman and {Datta}, Abhirup and {Mazumder}, Aishrila and {Majumdar}, Suman},
        title = "{Impact of calibration and position errors on astrophysical parameters of the HI 21cm signal}",
      journal = {\jcap},
         year = 2025,
        month = oct,
       volume = {2025},
       number = {10},
          eid = {035},
        pages = {035},
          doi = {10.1088/1475-7516/2025/10/035},
archivePrefix = {arXiv},
       eprint = {2502.20962},
 primaryClass = {astro-ph.CO},
       adsurl = {https://ui.adsabs.harvard.edu/abs/2025JCAP...10..035T}
}

@ARTICLE{2025JCAP...02..058P,
       author = {{Pal}, Samit Kumar and {Datta}, Abhirup and {Mazumder}, Aishrila},
        title = "{Ionospheric effect on the synthetic Epoch of Reionization observations with the SKA1-Low}",
      journal = {\jcap},
         year = 2025,
        month = feb,
       volume = {2025},
       number = {2},
          eid = {058},
        pages = {058},
          doi = {10.1088/1475-7516/2025/02/058},
       adsurl = {https://ui.adsabs.harvard.edu/abs/2025JCAP...02..058P}
}

@ARTICLE{2025arXiv251025886B,
       author = {{Beohar}, Eeshan and {Datta}, Abhirup and {Tripathi}, Anshuman and {Pal}, Samit Kumar and {Sagar}, Rashmi},
        title = "{Mitigating gain calibration errors from EoR observations with SKA1-Low AA*}",
      journal = {arXiv e-prints},
         year = 2025,
        month = oct,
          eid = {arXiv:2510.25886},
        pages = {arXiv:2510.25886},
archivePrefix = {arXiv},
       eprint = {2510.25886},
 primaryClass = {astro-ph.CO},
       adsurl = {https://ui.adsabs.harvard.edu/abs/2025arXiv251025886B}
}

@ARTICLE{2017JAI.....641011R,
       author = {{Reddy}, Suda Harshavardhan and {Kudale}, Sanjay and {Gokhale}, Upendra and {Halagalli}, Irappa and {Raskar}, Nilesh and {de}, Kishalay and {Gnanaraj}, Shelton and {Ajith Kumar}, B. and {Gupta}, Yashwant},
        title = "{A Wideband Digital Back-End for the Upgraded GMRT}",
      journal = {Journal of Astronomical Instrumentation},
         year = 2017,
        month = mar,
       volume = {6},
       number = {1},
          eid = {1641011-336},
        pages = {1641011-336},
          doi = {10.1142/S2251171716410117},
       adsurl = {https://ui.adsabs.harvard.edu/abs/2017JAI.....641011R}
}

@ARTICLE{2017CSci..113..707G,
       author = {{Gupta}, Y. and {Ajithkumar}, B. and {Kale}, H.~S. and {Nayak}, S. and {Sabhapathy}, S. and {Sureshkumar}, S. and {Swami}, R.~V. and {Chengalur}, J.~N. and {Ghosh}, S.~K. and {Ishwara-Chandra}, C.~H. and {Joshi}, B.~C. and {Kanekar}, N. and {Lal}, D.~V. and {Roy}, S.},
        title = "{The upgraded GMRT: opening new windows on the radio Universe}",
      journal = {Current Science},
         year = 2017,
        month = aug,
       volume = {113},
       number = {4},
        pages = {707-714},
          doi = {10.18520/cs/v113/i04/707-714},
       adsurl = {https://ui.adsabs.harvard.edu/abs/2017CSci..113..707G}
}

@ARTICLE{2014MNRAS.445.4351C,
       author = {{Choudhuri}, Samir and {Bharadwaj}, Somnath and {Ghosh}, Abhik and {Ali}, Sk. Saiyad},
        title = "{Visibility-based angular power spectrum estimation in low-frequency radio interferometric observations}",
      journal = {\mnras},
         year = 2014,
        month = dec,
       volume = {445},
       number = {4},
        pages = {4351-4365},
          doi = {10.1093/mnras/stu2027},
archivePrefix = {arXiv},
       eprint = {1409.7789},
 primaryClass = {astro-ph.CO},
       adsurl = {https://ui.adsabs.harvard.edu/abs/2014MNRAS.445.4351C}
}

@ARTICLE{2016MNRAS.463.4093C,
       author = {{Choudhuri}, Samir and {Bharadwaj}, Somnath and {Chatterjee}, Suman and {Ali}, Sk. Saiyad and {Roy}, Nirupam and {Ghosh}, Abhik},
        title = "{The visibility-based tapered gridded estimator (TGE) for the redshifted 21-cm power spectrum}",
      journal = {\mnras},
         year = 2016,
        month = dec,
       volume = {463},
       number = {4},
        pages = {4093-4107},
          doi = {10.1093/mnras/stw2254},
archivePrefix = {arXiv},
       eprint = {1609.01732},
 primaryClass = {astro-ph.CO},
       adsurl = {https://ui.adsabs.harvard.edu/abs/2016MNRAS.463.4093C}
}

@ARTICLE{2010ApJ...724..526D,
       author = {{Datta}, A. and {Bowman}, J.~D. and {Carilli}, C.~L.},
        title = "{Bright Source Subtraction Requirements for Redshifted 21 cm Measurements}",
      journal = {\apj},
         year = 2010,
        month = nov,
       volume = {724},
       number = {1},
        pages = {526-538},
          doi = {10.1088/0004-637X/724/1/526},
archivePrefix = {arXiv},
       eprint = {1005.4071},
 primaryClass = {astro-ph.CO},
       adsurl = {https://ui.adsabs.harvard.edu/abs/2010ApJ...724..526D}
}

@ARTICLE{2012ApJ...753...81P,
       author = {{Parsons}, Aaron and {Pober}, Jonathan and {McQuinn}, Matthew and {Jacobs}, Daniel and {Aguirre}, James},
        title = "{A Sensitivity and Array-configuration Study for Measuring the Power Spectrum of 21 cm Emission from Reionization}",
      journal = {\apj},
         year = 2012,
        month = jul,
       volume = {753},
       number = {1},
          eid = {81},
        pages = {81},
          doi = {10.1088/0004-637X/753/1/81},
archivePrefix = {arXiv},
       eprint = {1103.2135},
 primaryClass = {astro-ph.IM},
       adsurl = {https://ui.adsabs.harvard.edu/abs/2012ApJ...753...81P}
}

@ARTICLE{2012MNRAS.419.3491L,
       author = {{Liu}, Adrian and {Tegmark}, Max},
        title = "{How well can we measure and understand foregrounds with 21-cm experiments?}",
      journal = {\mnras},
         year = 2012,
        month = feb,
       volume = {419},
       number = {4},
        pages = {3491-3504},
          doi = {10.1111/j.1365-2966.2011.19989.x},
archivePrefix = {arXiv},
       eprint = {1106.0007},
 primaryClass = {astro-ph.CO},
       adsurl = {https://ui.adsabs.harvard.edu/abs/2012MNRAS.419.3491L}
}

@ARTICLE{2012MNRAS.423.2518C,
       author = {{Chapman}, Emma and {Abdalla}, Filipe B. and {Harker}, Geraint and {Jeli{\'c}}, Vibor and {Labropoulos}, Panagiotis and {Zaroubi}, Saleem and {Brentjens}, Michiel A. and {de Bruyn}, A.~G. and {Koopmans}, L.~V.~E.},
        title = "{Foreground removal using FASTICA: a showcase of LOFAR-EoR}",
      journal = {\mnras},
         year = 2012,
        month = jul,
       volume = {423},
       number = {3},
        pages = {2518-2532},
          doi = {10.1111/j.1365-2966.2012.21065.x},
archivePrefix = {arXiv},
       eprint = {1201.2190},
 primaryClass = {astro-ph.CO},
       adsurl = {https://ui.adsabs.harvard.edu/abs/2012MNRAS.423.2518C}
}

@ARTICLE{2020MNRAS.494.1936C,
       author = {{Choudhuri}, Samir and {Ghosh}, Abhik and {Roy}, Nirupam and {Bharadwaj}, Somnath and {Intema}, Huib T. and {Ali}, Sk Saiyad},
        title = "{All-sky angular power spectrum - I. Estimating brightness temperature fluctuations using the 150-MHz TGSS survey}",
      journal = {\mnras},
         year = 2020,
        month = may,
       volume = {494},
       number = {2},
        pages = {1936-1945},
          doi = {10.1093/mnras/staa762},
archivePrefix = {arXiv},
       eprint = {2003.07869},
 primaryClass = {astro-ph.CO},
       adsurl = {https://ui.adsabs.harvard.edu/abs/2020MNRAS.494.1936C}
}

@ARTICLE{2023JApA...44...37B,
       author = {{Buch}, Kaushal D. and {Kale}, Ruta and {Muley}, Mekhala and {Kudale}, Sanjay and {Ajithkumar}, B.},
        title = "{Real-time RFI filtering for uGMRT: Overview of the released system and relevance to the SKA}",
      journal = {Journal of Astrophysics and Astronomy},
         year = 2023,
        month = jun,
       volume = {44},
       number = {1},
          eid = {37},
        pages = {37},
          doi = {10.1007/s12036-023-09919-x},
archivePrefix = {arXiv},
       eprint = {2301.07402},
 primaryClass = {astro-ph.IM},
       adsurl = {https://ui.adsabs.harvard.edu/abs/2023JApA...44...37B}
}

@ARTICLE{2021A&A...648A...2S,
       author = {{Sabater}, J. and {Best}, P.~N. and {Tasse}, C. and {Hardcastle}, M.~J. and {Shimwell}, T.~W. and {Nisbet}, D. and {Jelic}, V. and {Callingham}, J.~R. and {R{\"o}ttgering}, H.~J.~A. and {Bonato}, M. and {Bondi}, M. and {Ciardi}, B. and {Cochrane}, R.~K. and {Jarvis}, M.~J. and {Kondapally}, R. and {Koopmans}, L.~V.~E. and {O'Sullivan}, S.~P. and {Prandoni}, I. and {Schwarz}, D.~J. and {Smith}, D.~J.~B. and {Wang}, L. and {Williams}, W.~L. and {Zaroubi}, S.},
        title = "{The LOFAR Two-meter Sky Survey: Deep Fields Data Release 1. II. The ELAIS-N1 LOFAR deep field}",
      journal = {\aap},
         year = 2021,
        month = apr,
       volume = {648},
          eid = {A2},
        pages = {A2},
          doi = {10.1051/0004-6361/202038828},
archivePrefix = {arXiv},
       eprint = {2011.08211},
 primaryClass = {astro-ph.GA},
       adsurl = {https://ui.adsabs.harvard.edu/abs/2021A&A...648A...2S}
}

@ARTICLE{2024JCAP...05..068G,
       author = {{Gayen}, Saikat and {Sagar}, Rashmi and {Mangla}, Sarvesh and {Dutta}, Prasun and {Roy}, Nirupam and {Chakraborty}, Arnab and {Kumar}, Jais and {Datta}, Abhirup and {Choudhuri}, Samir},
        title = "{Calibration requirement for Epoch of Reionization 21-cm signal observation. Part III. Bias and variance in uGMRT ELAIS-N1 field power spectrum}",
      journal = {\jcap},
         year = 2024,
        month = may,
       volume = {2024},
       number = {5},
          eid = {068},
        pages = {068},
          doi = {10.1088/1475-7516/2024/05/068},
archivePrefix = {arXiv},
       eprint = {2402.18306},
 primaryClass = {astro-ph.IM},
       adsurl = {https://ui.adsabs.harvard.edu/abs/2024JCAP...05..068G}
}

@ARTICLE{2025JCAP...07..024G,
       author = {{Gayen}, Saikat and {Kumar}, Jais and {Dutta}, Prasun and {Elahi}, Khandakar Md Asif and {Choudhuri}, Samir and {Roy}, Nirupam},
        title = "{Calibration requirements for epoch of reionization 21-cm signal observations. Part IV. Bias and variance with time and frequency correlated residual gains}",
      journal = {\jcap},
         year = 2025,
        month = jul,
       volume = {2025},
       number = {7},
          eid = {024},
        pages = {024},
          doi = {10.1088/1475-7516/2025/07/024},
archivePrefix = {arXiv},
       eprint = {2503.23825},
 primaryClass = {astro-ph.CO},
       adsurl = {https://ui.adsabs.harvard.edu/abs/2025JCAP...07..024G}
}

@ARTICLE{2025arXiv251102375S,
       author = {{Sagar}, Rashmi and {Datta}, Abhirup and {Chakraborty}, Arnab and {Roy}, Nirupam and {Sinha}, Akriti and {Mazumder}, Aishrila and {Dutta}, Prasun and {Elahi}, Kh. Md. Asif and {Datta}, Kanan K. and {Choudhuri}, Samir and {Bharadwaj}, Somnath and {Pal}, Srijita and {Tripathi}, Anshuman and {Majumdar}, Suman and {Choudhury}, Tirthankar Roy and {Saiyad Ali}, Sk.},
        title = "{ELAIS-N1 Deep Field uGMRT Band-2: Constraints on Diffuse Galactic Synchrotron Emission Power Spectrum}",
      journal = {arXiv e-prints},
         year = 2025,
        month = nov,
          eid = {arXiv:2511.02375},
        pages = {arXiv:2511.02375},
          doi = {10.48550/arXiv.2511.02375},
archivePrefix = {arXiv},
       eprint = {2511.02375},
 primaryClass = {astro-ph.HE},
       adsurl = {https://ui.adsabs.harvard.edu/abs/2025arXiv251102375S}
}

@ARTICLE{2025MNRAS.544..321E,
       author = {{Elahi}, Khandakar Md Asif and {Choudhuri}, Samir and {Roy}, Nirupam and {Rashid}, Md and {Bull}, Philip and {Lal}, Dharam Vir},
        title = "{Deep uGMRT observations for enhanced calibration of 21 cm arrays {\textendash} I. First image and source catalogue}",
      journal = {\mnras},
         year = 2025,
        month = nov,
       volume = {544},
       number = {1},
        pages = {321-342},
          doi = {10.1093/mnras/staf1764},
       adsurl = {https://ui.adsabs.harvard.edu/abs/2025MNRAS.544..321E}
}

@ARTICLE{2014MNRAS.444..606O,
       author = {{Offringa}, A.~R. and {McKinley}, B. and {Hurley-Walker}, N. and {Briggs}, F.~H. and {Wayth}, R.~B. and {Kaplan}, D.~L. and {Bell}, M.~E. and {Feng}, L. and {Neben}, A.~R. and {Hughes}, J.~D. and {Rhee}, J. and {Murphy}, T. and {Bhat}, N.~D.~R. and {Bernardi}, G. and {Bowman}, J.~D. and {Cappallo}, R.~J. and {Corey}, B.~E. and {Deshpande}, A.~A. and {Emrich}, D. and {Ewall-Wice}, A. and {Gaensler}, B.~M. and {Goeke}, R. and {Greenhill}, L.~J. and {Hazelton}, B.~J. and {Hindson}, L. and {Johnston-Hollitt}, M. and {Jacobs}, D.~C. and {Kasper}, J.~C. and {Kratzenberg}, E. and {Lenc}, E. and {Lonsdale}, C.~J. and {Lynch}, M.~J. and {McWhirter}, S.~R. and {Mitchell}, D.~A. and {Morales}, M.~F. and {Morgan}, E. and {Kudryavtseva}, N. and {Oberoi}, D. and {Ord}, S.~M. and {Pindor}, B. and {Procopio}, P. and {Prabu}, T. and {Riding}, J. and {Roshi}, D.~A. and {Shankar}, N. Udaya and {Srivani}, K.~S. and {Subrahmanyan}, R. and {Tingay}, S.~J. and {Waterson}, M. and {Webster}, R.~L. and {Whitney}, A.~R. and {Williams}, A. and {Williams}, C.~L.},
        title = "{WSCLEAN: an implementation of a fast, generic wide-field imager for radio astronomy}",
      journal = {\mnras},
         year = 2014,
        month = oct,
       volume = {444},
       number = {1},
        pages = {606-619},
          doi = {10.1093/mnras/stu1368},
archivePrefix = {arXiv},
       eprint = {1407.1943},
 primaryClass = {astro-ph.IM},
       adsurl = {https://ui.adsabs.harvard.edu/abs/2014MNRAS.444..606O}
}

@article{barry_calibration_2016,
	title = {Calibration requirements for detecting the 21 cm epoch of reionization power spectrum and implications for the {SKA}},
	volume = {461},
	issn = {0035-8711},
	url = {https://academic.oup.com/mnras/article/461/3/3135/2608487/Calibration-requirements-for-detecting-the-21-cm},
	doi = {10.1093/mnras/stw1380},
	number = {3},
	urldate = {2017-02-01},
	journal = {Monthly Notices of the Royal Astronomical Society},
	author = {Barry, N. and Hazelton, B. and Sullivan, I. and Morales, M. F. and Pober, J. C.},
	month = sep,
	year = {2016},
	pages = {3135--3144},
}

@article{barry_improving_2019,
	title = {Improving the {Epoch} of {Reionization} {Power} {Spectrum} {Results} from {Murchison} {Widefield} {Array} {Season} 1 {Observations}},
	volume = {884},
	issn = {0004-637X},
	url = {https://doi.org/10.3847%2F1538-4357%2Fab40a8},
	doi = {10.3847/1538-4357/ab40a8},
	language = {en},
	number = {1},
	urldate = {2019-11-20},
	journal = {The Astrophysical Journal},
	author = {Barry, N. and Wilensky, M. and Trott, C. M. and Pindor, B. and Beardsley, A. P. and Hazelton, B. J. and Sullivan, I. S. and Morales, M. F. and Pober, J. C. and Line, J. and Greig, B. and Byrne, R. and Lanman, A. and Li, W. and Jordan, C. H. and Joseph, R. C. and McKinley, B. and Rahimi, M. and Yoshiura, S. and Bowman, J. D. and Gaensler, B. M. and Hewitt, J. N. and Jacobs, D. C. and Mitchell, D. A. and Shankar, N. Udaya and Sethi, S. K. and Subrahmanyan, R. and Tingay, S. J. and Webster, R. L. and Wyithe, J. S. B.},
	month = oct,
	year = {2019},
	pages = {1},
}

@article{beardsley_first_2016,
	title = {First {Season} {MWA} {EoR} {Power} spectrum {Results} at {Redshift} 7},
	volume = {833},
	issn = {0004-637X},
	url = {http://stacks.iop.org/0004-637X/833/i=1/a=102},
	doi = {10.3847/1538-4357/833/1/102},
	language = {en},
	number = {1},
	urldate = {2017-04-19},
	journal = {The Astrophysical Journal},
	author = {Beardsley, A. P. and Hazelton, B. J. and Sullivan, I. S. and Carroll, P. and Barry, N. and Rahimi, M. and Pindor, B. and Trott, C. M. and {J. Line} and Jacobs, Daniel C. and Morales, M. F. and Pober, J. C. and Bernardi, G. and Bowman, Judd D. and Busch, M. P. and Briggs, F. and Cappallo, R. J. and Corey, B. E. and Oliveira-Costa, A. de and Dillon, Joshua S. and Emrich, D. and Ewall-Wice, A. and Feng, L. and Gaensler, B. M. and Goeke, R. and Greenhill, L. J. and Hewitt, J. N. and Hurley-Walker, N. and Johnston-Hollitt, M. and Kaplan, D. L. and Kasper, J. C. and Kim, H. S. and Kratzenberg, E. and Lenc, E. and Loeb, A. and Lonsdale, C. J. and Lynch, M. J. and McKinley, B. and McWhirter, S. R. and Mitchell, D. A. and Morgan, E. and Neben, A. R. and Thyagarajan, Nithyanandan and Oberoi, D. and Offringa, A. R. and Ord, S. M. and Paul, S. and Prabu, T. and Procopio, P. and Riding, J. and Rogers, A. E. E. and Roshi, A. and Shankar, N. Udaya and Sethi, Shiv K. and Srivani, K. S. and Subrahmanyan, R. and Tegmark, M. and Tingay, S. J. and Waterson, M. and Wayth, R. B. and Webster, R. L. and Whitney, A. R. and Williams, A. and Williams, C. L. and Wu, C. and Wyithe, J. S. B.},
	year = {2016},
	pages = {102},
}

@article{byrne_fundamental_2019,
	title = {Fundamental {Limitations} on the {Calibration} of {Redundant} 21 cm {Cosmology} {Instruments} and {Implications} for {HERA} and the {SKA}},
	volume = {875},
	issn = {0004-637X},
	url = {https://doi.org/10.3847%2F1538-4357%2Fab107d},
	doi = {10.3847/1538-4357/ab107d},
	language = {en},
	number = {1},
	urldate = {2019-04-18},
	journal = {The Astrophysical Journal},
	author = {Byrne, Ruby and Morales, Miguel F. and Hazelton, Bryna and Li, Wenyang and Barry, Nichole and Beardsley, Adam P. and Joseph, Ronniy and Pober, Jonathan and Sullivan, Ian and Trott, Cathryn},
	month = apr,
	year = {2019},
	pages = {70},
}

@article{cheng_characterizing_2018,
	title = {Characterizing {Signal} {Loss} in the 21 cm {Reionization} {Power} {Spectrum}: {A} {Revised} {Study} of {PAPER}-64},
	volume = {868},
	issn = {0004-637X},
	shorttitle = {Characterizing {Signal} {Loss} in the 21 cm {Reionization} {Power} {Spectrum}},
	url = {https://doi.org/10.3847/1538-4357/aae833},
	doi = {10.3847/1538-4357/aae833},
	language = {en},
	number = {1},
	urldate = {2025-10-13},
	journal = {The Astrophysical Journal},
	author = {Cheng, Carina and Parsons, Aaron R. and Kolopanis, Matthew and Jacobs, Daniel C. and Liu, Adrian and Kohn, Saul A. and Aguirre, James E. and Pober, Jonathan C. and Ali, Zaki S. and Bernardi, Gianni and Bradley, Richard F. and Carilli, Chris L. and DeBoer, David R. and Dexter, Matthew R. and Dillon, Joshua S. and Klima, Pat and MacMahon, David H. E. and Moore, David F. and Nunhokee, Chuneeta D. and Walbrugh, William P. and Walker, Andre},
	month = nov,
	year = {2018},
	pages = {26},
}

@article{line_verifying_2024,
	title = {Verifying the {Australian} {MWA} {EoR} pipeline {I}: 21-cm sky model and correlated measurement density},
	issn = {1323-3580, 1448-6083},
	shorttitle = {Verifying the {Australian} {MWA} {EoR} pipeline {I}},
	url = {https://www.cambridge.org/core/journals/publications-of-the-astronomical-society-of-australia/article/abs/verifying-the-australian-mwa-eor-pipeline-i-21cm-sky-model-and-correlated-measurement-density/61AEF74A93A547D2BF13C0CC8A8FFB59},
	doi = {10.1017/pasa.2024.31},
	language = {en},
	urldate = {2024-06-06},
	journal = {Publications of the Astronomical Society of Australia},
	author = {Line, J. L. B. and Trott, C. M. and Cook, J. H. and Greig, B. and Barry, N. and Jordan, C. H.},
	month = apr,
	year = {2024},
	pages = {1--13},
}

@article{line_woden_2022,
	title = {`{WODEN}`: {A} {CUDA}-enabled package to simulate low-frequency radio interferometric data},
	volume = {7},
	issn = {2475-9066},
	shorttitle = {`{WODEN}`},
	url = {https://joss.theoj.org/papers/10.21105/joss.03676},
	doi = {10.21105/joss.03676},
	language = {en},
	number = {69},
	urldate = {2023-02-07},
	journal = {Journal of Open Source Software},
	author = {Line, Jack L. b},
	month = jan,
	year = {2022},
	pages = {3676},
}

@article{shimwell_lofar_2017,
	title = {The {LOFAR} {Two}-metre {Sky} {Survey} - {I}. {Survey} description and preliminary data release},
	volume = {598},
	copyright = {© ESO, 2017},
	issn = {0004-6361, 1432-0746},
	url = {https://www.aanda.org/articles/aa/abs/2017/02/aa29313-16/aa29313-16.html},
	doi = {10.1051/0004-6361/201629313},
	language = {en},
	urldate = {2025-10-30},
	journal = {Astronomy \& Astrophysics},
	author = {Shimwell, T. W. and Röttgering, H. J. A. and Best, P. N. and Williams, W. L. and Dijkema, T. J. and Gasperin, F. de and Hardcastle, M. J. and Heald, G. H. and Hoang, D. N. and Horneffer, A. and Intema, H. and Mahony, E. K. and Mandal, S. and Mechev, A. P. and Morabito, L. and Oonk, J. B. R. and Rafferty, D. and Retana-Montenegro, E. and Sabater, J. and Tasse, C. and Weeren, R. J. van and Brüggen, M. and Brunetti, G. and Chyży, K. T. and Conway, J. E. and Haverkorn, M. and Jackson, N. and Jarvis, M. J. and McKean, J. P. and Miley, G. K. and Morganti, R. and White, G. J. and Wise, M. W. and Bemmel, I. M. van and Beck, R. and Brienza, M. and Bonafede, A. and Rivera, G. Calistro and Cassano, R. and Clarke, A. O. and Cseh, D. and Deller, A. and Drabent, A. and Driel, W. van and Engels, D. and Falcke, H. and Ferrari, C. and Fröhlich, S. and Garrett, M. A. and Harwood, J. J. and Heesen, V. and Hoeft, M. and Horellou, C. and Israel, F. P. and Kapińska, A. D. and Kunert-Bajraszewska, M. and McKay, D. J. and Mohan, N. R. and Orrú, E. and Pizzo, R. F. and Prandoni, I. and Schwarz, D. J. and Shulevski, A. and Sipior, M. and Smith, D. J. B. and Sridhar, S. S. and Steinmetz, M. and Stroe, A. and Varenius, E. and Werf, P. P. van der and Zensus, J. A. and Zwart, J. T. L.},
	month = feb,
	year = {2017},
	pages = {A104},
}

@article{chokshi_necessity_2024,
	title = {The necessity of individually validated beam models for an interferometric epoch of reionization detection},
	volume = {534},
	issn = {0035-8711},
	url = {https://dx.doi.org/10.1093/mnras/stae2264},
	doi = {10.1093/mnras/stae2264},
	language = {en},
	number = {3},
	urldate = {2024-11-13},
	journal = {Monthly Notices of the Royal Astronomical Society},
	author = {Chokshi, A. and Barry, N. and Line, J. L. B. and Jordan, C. H. and Pindor, B. and Webster, R. L.},
	month = oct,
	year = {2024},
	pages = {2475--2484},
}

@article{joseph_calibration_2020,
	title = {Calibration and 21-cm power spectrum estimation in the presence of antenna beam variations},
	volume = {492},
	issn = {0035-8711},
	url = {https://doi.org/10.1093/mnras/stz3375},
	doi = {10.1093/mnras/stz3375},
	number = {2},
	urldate = {2023-11-15},
	journal = {Monthly Notices of the Royal Astronomical Society},
	author = {Joseph, Ronniy C and Trott, C M and Wayth, R B and Nasirudin, A},
	month = feb,
	year = {2020},
	pages = {2017--2028},
}

@article{hurley-walker_galactic_2017,
	title = {{GaLactic} and {Extragalactic} {All}-sky {Murchison} {Widefield} {Array} ({GLEAM}) survey – {I}. {A} low-frequency extragalactic catalogue},
	volume = {464},
	issn = {0035-8711},
	url = {https://doi.org/10.1093/mnras/stw2337},
	doi = {10.1093/mnras/stw2337},
	number = {1},
	urldate = {2025-10-30},
	journal = {Monthly Notices of the Royal Astronomical Society},
	author = {Hurley-Walker, N and Callingham, J R and Hancock, P J and Franzen, T M O and Hindson, L and Kapińska, A D and Morgan, J and Offringa, A R and Wayth, R B and Wu, C and Zheng, Q and Murphy, T and Bell, M E and Dwarakanath, K S and For, B and Gaensler, B M and Johnston-Hollitt, M and Lenc, E and Procopio, P and Staveley-Smith, L and Ekers, R and Bowman, J D and Briggs, F and Cappallo, R J and Deshpande, A A and Greenhill, L and Hazelton, B J and Kaplan, D L and Lonsdale, C J and McWhirter, S R and Mitchell, D A and Morales, M F and Morgan, E and Oberoi, D and Ord, S M and Prabu, T and Shankar, N Udaya and Srivani, K S and Subrahmanyan, R and Tingay, S J and Webster, R L and Williams, A and Williams, C L},
	month = jan,
	year = {2017},
	pages = {1146--1167},
}

@article{lanman_pyuvsim_2019,
	title = {pyuvsim: {A} comprehensive simulation package for radio interferometers in python},
	volume = {4},
	issn = {2475-9066},
	shorttitle = {pyuvsim},
	url = {https://joss.theoj.org/papers/10.21105/joss.01234},
	doi = {10.21105/joss.01234},
	language = {en},
	number = {37},
	urldate = {2025-10-30},
	journal = {Journal of Open Source Software},
	author = {Lanman, Adam E. and Hazelton, Bryna J. and Jacobs, Daniel C. and Kolopanis, Matthew J. and Pober, Jonathan C. and Aguirre, James E. and Thyagarajan, Nithyanandan},
	month = may,
	year = {2019},
	pages = {1234},
}

@article{dijkema_dp3_2023,
	title = {{DP3}: {Streaming} processing pipeline for radio interferometric data},
	shorttitle = {{DP3}},
	url = {https://ui.adsabs.harvard.edu/abs/2023ascl.soft05014D},
	urldate = {2025-10-30},
	journal = {Astrophysics Source Code Library},
	author = {Dijkema, T. J. and Nijhuis, M. and van Diepen, G. and Offringa, A. and Krombeen, L. and de Wever, M. and Maljaars, J. and Loose, M.},
	month = may,
	year = {2023},
	note = {ADS Bibcode: 2023ascl.soft05014D},
	pages = {ascl:2305.014},
}

@ARTICLE{Cox_fftvis_2025,
       author = {{Cox}, Tyler A. and {Murray}, Steven G. and {Parsons}, Aaron R. and {Dillon}, Joshua S. and {Mandar}, Kartik and {Martinot}, Zachary E. and {Pascua}, Robert and {Kittiwisit}, Piyanat and {Aguirre}, James E.},
        title = "{fftvis: A Non-Uniform Fast Fourier Transform Based Interferometric Visibility Simulator}",
      journal = {arXiv e-prints},
         year = 2025,
        month = jun,
          eid = {arXiv:2506.02130},
        pages = {arXiv:2506.02130},
          doi = {10.48550/arXiv.2506.02130},
archivePrefix = {arXiv},
       eprint = {2506.02130},
 primaryClass = {astro-ph.IM},
       adsurl = {https://ui.adsabs.harvard.edu/abs/2025arXiv250602130C}
}

@ARTICLE{Kittiwisit_matvis_2024,
       author = {{Kittiwisit}, Piyanat and {Murray}, Steven G. and {Garsden}, Hugh and {Bull}, Philip and {Wilensky}, Michael J. and {Cain}, Christopher and {Parsons}, Aaron R. and {Sipple}, Jackson and {Adams}, Tyrone and {Aguirre}, James E. and {Baartman}, Rushelle and {Beardsley}, Adam P. and {Berkhout}, Lindsay M. and {Bernardi}, Gianni and {Billings}, Tashalee S. and {Bowman}, Judd D. and {Bradley}, Richard F. and {Burba}, Jacob and {Carey}, Steven and {Carilli}, Chris L. and {Chen}, Kai-Feng and {Choudhuri}, Samir and {Cox}, Tyler and {DeBoer}, David R. and {Acedo}, Eloy de Lera and {Dexter}, Matt and {Dillon}, Joshua S. and {Eksteen}, Nico and {Ely}, John and {Ewall-Wice}, Aaron and {Fagnoni}, Nicolas and {Furlanetto}, Steven R. and {Gale-Sides}, Kingsley and {Gehlot}, Bharat Kumar and {Glendenning}, Brian and {Gorce}, Adelie and {Gorthi}, Deepthi and {Greig}, Bradley and {Grobbelaar}, Jasper and {Halday}, Ziyaad and {Hazelton}, Bryna J. and {Hewitt}, Jacqueline N. and {Hickish}, Jack and {Jacobs}, Daniel C. and {Josaitis}, Alec and {Kern}, Nicholas S. and {Kerrigan}, Joshua and {Kim}, Honggeun and {Kolopanis}, Matthew and {Lanman}, Adam and {Plante}, Paul La and {Liu}, Adrian and {Ma}, Yin-Zhe and {MacMahon}, David H.~E. and {Malan}, Lourence and {Malgas}, Cresshim and {Malgas}, Keith and {Marero}, Bradley and {Martinot}, Zachary E. and {McBride}, Lisa and {Mesinger}, Andrei and {Molewa}, Mathakane and {Morales}, Miguel F. and {Mosiane}, Tshegofalang and {Nunhokee}, Chuneeta Devi and {Nuwegeld}, Hans and {Pascua}, Robert and {Qin}, Yuxiang and {Rath}, Eleanor and {Razavi-Ghods}, Nima and {Robnett}, James and {Santos}, Mario G. and {Sims}, Peter and {Singh}, Saurabh and {Storer}, Dara and {Swarts}, Hilton and {Tan}, Jianrong and {Thyagarajan}, Nithyanandan and {van Wyngaarden}, Pieter and {Xu}, Zhilei and {Zheng}, Haoxuan},
        title = "{matvis: a matrix-based visibility simulator for fast forward modelling of many-element 21 cm arrays}",
      journal = {RAS Techniques and Instruments},
         year = 2025,
        month = jan,
       volume = {4},
          eid = {rzaf001},
        pages = {rzaf001},
          doi = {10.1093/rasti/rzaf001},
archivePrefix = {arXiv},
       eprint = {2312.09763},
 primaryClass = {astro-ph.IM},
       adsurl = {https://ui.adsabs.harvard.edu/abs/2025RASTI...4....1K}
}

@ARTICLE{NCP2013,
   author = {{Yatawatta}, S. and {de Bruyn}, A.~G. and {Brentjens}, M.~A. and
  {Labropoulos}, P.},
    title = "{Initial deep LOFAR observations of epoch of reionization windows. I. The north celestial pole}",
  journal = {Astronomy \& Astrophysics},
     year = 2013,
    month = feb,
   volume = 550,
      eid = {A136},
    pages = {A136},
}

@ARTICLE{Patil2017,
   author = {{Patil}, A.~H. and {Yatawatta}, S. and {Koopmans}, L.~V.~E. and others},
    title = "{Upper Limits on the 21 cm Epoch of Reionization Power Spectrum from One Night with LOFAR}",
  journal = {\apj},
archivePrefix = "arXiv",
   eprint = {1702.08679},
     year = 2017,
    month = mar,
   volume = 838,
      eid = {65},
    pages = {65},
      doi = {10.3847/1538-4357/aa63e7},
}

@article{mertens2020,
  title={Improved upper limits on the 21 cm signal power spectrum of neutral hydrogen at z 9.1 from {LOFAR}},
  author={Mertens, FG and Mevius, M and Koopmans, LVE and Offringa, AR and Mellema, Garrelt and Zaroubi, S and Brentjens, MA and Gan, H and Gehlot, BK and Pandey, VN and others},
  journal={Monthly Notices of the Royal Astronomical Society},
  volume={493},
  number={2},
  pages={1662--1685},
  year={2020},
  publisher={Oxford University Press}
}

@article{mertens2025,
  title={Deeper multi-redshift upper limits on the epoch of reionization 21 cm signal power spectrum from {LOFAR} between z=8.3 and z=10.1},
  author={Mertens, FG and others},
  journal={\aap},
  year={2025},
  month = jun,
  volume = {698},
  eid = {A186},
  pages = {A186},
  doi = {10.1051/0004-6361/202554158},
}

@article{munshi2024first,
  title={First upper limits on the 21 cm signal power spectrum from cosmic dawn from one night of observations with NenuFAR},
  author={Munshi, S and Mertens, FG and Koopmans, LVE and Offringa, AR and Semelin, B and Aubert, D and Barkana, R and Bracco, A and Brackenhoff, SA and Cecconi, B and others},
  journal={Astronomy \& Astrophysics},
  volume={681},
  pages={A62},
  year={2024},
  publisher={EDP Sciences}
}

@article{munshi2025mitigating,
  title={Mitigating incoherent excess variance in high-redshift 21-cm observations with multi-output cross Gaussian process regression},
  author={Munshi, S and Koopmans, LVE and Mertens, FG and Offringa, AR and Brackenhoff, SA and Ceccotti, E and Chege, JK and Gao, LY and Ghosh, S and Mevius, M and others},
  journal={arXiv preprint arXiv:2508.08235},
  year={2025}
}

@article{munshi2025near,
  title={Near-field imaging of local interference in radio interferometric data-Impact on the redshifted 21 cm power spectrum},
  author={Munshi, S and Mertens, FG and Koopmans, LVE and Mevius, M and Offringa, AR and Semelin, B and Viou, C and Bracco, A and Brackenhoff, SA and Ceccotti, E and others},
  journal={Astronomy \& Astrophysics},
  volume={697},
  pages={A203},
  year={2025},
  publisher={EDP Sciences}
}

@article{mevius2021,
  title={A numerical study of 21-cm signal suppression and noise increase in
     direction-dependent calibration of {LOFAR} data},
  author={Mevius, M and others},
  journal = {Monthly Notices of the Royal Astronomical Society},
  volume = {509},
  number = {3},
  pages = {3693-3702},
  year = {2021},
  month = {11},
}

@ARTICLE{Patil2016,
   author = {{Patil}, A.~H. and {Yatawatta}, S. and {Zaroubi}, S. and {Koopmans}, L.~V.~E. and
  {de Bruyn}, A.~G. and {Jeli{\'c}}, V. and {Ciardi}, B. and {Iliev}, I.~T. and
  {Mevius}, M. and {Pandey}, V.~N. and {Gehlot}, B.~K.},
    title = "{Systematic biases in low-frequency radio interferometric data due to calibration: the LOFAR-EoR case}",
  journal = {\mnras},
archivePrefix = "arXiv",
   eprint = {1605.07619},
 primaryClass = "astro-ph.IM",
     year = 2016,
    month = dec,
   volume = 463,
    pages = {4317-4330},
      doi = {10.1093/mnras/stw2277},
   adsurl = {http://adsabs.harvard.edu/abs/2016MNRAS.463.4317P}
}

@ARTICLE{Millad2018,
   author = {{Mouri Sardarabadi}, A. and {Koopmans}, L.~V.~E.},
    title = "{Quantifying Suppression of the Cosmological 21-cm Signal due to Direction Dependent Gain Calibration in Radio Interferometers}",
  journal = {ArXiv e-prints},
archivePrefix = "arXiv",
   eprint = {1809.03755},
 primaryClass = "astro-ph.IM",
     year = 2018,
    month = sep,
   adsurl = {http://adsabs.harvard.edu/abs/2018arXiv180903755M}
}

@article{DCAL,
author = {{Yatawatta}, S.},
title = {Distributed radio interferometric calibration},
volume = {449},
number = {4},
pages = {4506-4514},
year = {2015},
doi = {10.1093/mnras/stv596},
journal = {\mnras},
}

@article{Cees2023,
author = {F. Di Vruno and B. Winkel and C. G. Bassa and G. I. G. Jozsa and M. A. Brentjens and A. Jessner and S. Garrington},
title = {Unintended electromagnetic radiation from Starlink satellites detected with {LOFAR} between 110 and 188 {MHz}},
journal={Astronomy and Astrophysics},
month = jul,
eid = {arXiv:2307.02316},
pages = {arXiv:2307.02316},
doi = {10.48550/arXiv.2307.02316},
volume = {1},
number = {1},
year = {2023},
URL = {
https://doi.org/10.1051/0004-6361/202346374
},
}

@ARTICLE{Mertens2024,
       author = {{Mertens}, Florent G. and {Bobin}, J{\'e}r{\^o}me and {Carucci}, Isabella P.},
        title = "{Retrieving the 21-cm signal from the Epoch of Reionization with learnt Gaussian process kernels}",
      journal = {\mnras},
         year = 2024,
        month = jan,
       volume = {527},
       number = {2},
        pages = {3517-3531},
          doi = {10.1093/mnras/stad3430},
}

@ARTICLE{YA2021MNRAS,
       author = {{Yatawatta}, Sarod and {Avruch}, Ian M.},
        title = "{Deep reinforcement learning for smart calibration of radio telescopes}",
      journal = {\mnras},
         year = 2021,
        month = aug,
       volume = {505},
       number = {2},
        pages = {2141-2150},
          doi = {10.1093/mnras/stab1401},
}

@article{Kern2019,
doi = {10.3847/1538-4357/ab3e73},
url = {https://doi.org/10.3847/1538-4357/ab3e73},
year = {2019},
month = {oct},
publisher = {The American Astronomical Society},
volume = {884},
number = {2},
pages = {105},
author = {Kern, Nicholas S. and Parsons, Aaron R. and Dillon, Joshua S. and Lanman, Adam E. and Fagnoni, Nicolas and de Lera Acedo, Eloy},
title = {Mitigating Internal Instrument Coupling for 21 cm Cosmology. I. Temporal and Spectral Modeling in Simulations},
journal = {The Astrophysical Journal},
}

@ARTICLE{Kern2020a,
       author = {{Kern}, Nicholas S. and {Parsons}, Aaron R. and {Dillon}, Joshua S. and {Lanman}, Adam E. and {Liu}, Adrian and {Bull}, Philip and {Ewall-Wice}, Aaron and {Abdurashidova}, Zara and {Aguirre}, James E. and {Alexander}, Paul and {Ali}, Zaki S. and {Balfour}, Yanga and {Beardsley}, Adam P. and {Bernardi}, Gianni and {Bowman}, Judd D. and {Bradley}, Richard F. and {Burba}, Jacob and {Carilli}, Chris L. and {Cheng}, Carina and {DeBoer}, David R. and {Dexter}, Matt and {de Lera Acedo}, Eloy and {Fagnoni}, Nicolas and {Fritz}, Randall and {Furlanetto}, Steve R. and {Glendenning}, Brian and {Gorthi}, Deepthi and {Greig}, Bradley and {Grobbelaar}, Jasper and {Halday}, Ziyaad and {Hazelton}, Bryna J. and {Hewitt}, Jacqueline N. and {Hickish}, Jack and {Jacobs}, Daniel C. and {Julius}, Austin and {Kerrigan}, Joshua and {Kittiwisit}, Piyanat and {Kohn}, Saul A. and {Kolopanis}, Matthew and {La Plante}, Paul and {Lekalake}, Telalo and {MacMahon}, David and {Malan}, Lourence and {Malgas}, Cresshim and {Maree}, Matthys and {Martinot}, Zachary E. and {Matsetela}, Eunice and {Mesinger}, Andrei and {Molewa}, Mathakane and {Morales}, Miguel F. and {Mosiane}, Tshegofalang and {Murray}, Steven G. and {Neben}, Abraham R. and {Parsons}, Aaron R. and {Patra}, Nipanjana and {Pieterse}, Samantha and {Pober}, Jonathan C. and {Razavi-Ghods}, Nima and {Ringuette}, Jon and {Robnett}, James and {Rosie}, Kathryn and {Sims}, Peter and {Smith}, Craig and {Syce}, Angelo and {Thyagarajan}, Nithyanandan and {Williams}, Peter K.~G. and {Zheng}, Haoxuan},
        title = "{Mitigating Internal Instrument Coupling for 21 cm Cosmology. II. A Method Demonstration with the Hydrogen Epoch of Reionization Array}",
      journal = {\apj},
         year = 2020,
        month = jan,
       volume = {888},
       number = {2},
          eid = {70},
        pages = {70},
          doi = {10.3847/1538-4357/ab5e8a},
archivePrefix = {arXiv},
       eprint = {1909.11733},
 primaryClass = {astro-ph.IM},
       adsurl = {https://ui.adsabs.harvard.edu/abs/2020ApJ...888...70K}
}

@ARTICLE{Kern2020b,
       author = {{Kern}, Nicholas S. and {Dillon}, Joshua S. and {Parsons}, Aaron R. and {Carilli}, Christopher L. and {Bernardi}, Gianni and {Abdurashidova}, Zara and {Aguirre}, James E. and {Alexander}, Paul and {Ali}, Zaki S. and {Balfour}, Yanga and {Beardsley}, Adam P. and {Billings}, Tashalee S. and {Bowman}, Judd D. and {Bradley}, Richard F. and {Bull}, Philip and {Burba}, Jacob and {Carey}, Steven and {Cheng}, Carina and {DeBoer}, David R. and {Dexter}, Matt and {de Lera Acedo}, Eloy and {Ely}, John and {Ewall-Wice}, Aaron and {Fagnoni}, Nicolas and {Fritz}, Randall and {Furlanetto}, Steve R. and {Gale-Sides}, Kingsley and {Glendenning}, Brian and {Gorthi}, Deepthi and {Greig}, Bradley and {Grobbelaar}, Jasper and {Halday}, Ziyaad and {Hazelton}, Bryna J. and {Hewitt}, Jacqueline N. and {Hickish}, Jack and {Jacobs}, Daniel C. and {Julius}, Austin and {Kerrigan}, Joshua and {Kittiwisit}, Piyanat and {Kohn}, Saul A. and {Kolopanis}, Matthew and {Lanman}, Adam and {La Plante}, Paul and {Lekalake}, Telalo and {Liu}, Adrian and {MacMahon}, David and {Malan}, Lourence and {Malgas}, Cresshim and {Maree}, Matthys and {Martinot}, Zachary E. and {Matsetela}, Eunice and {Mesinger}, Andrei and {Molewa}, Mathakane and {Morales}, Miguel F. and {Mosiane}, Tshegofalang and {Murray}, Steven G. and {Neben}, Abraham R. and {Nikolic}, Bojan and {Nunhokee}, Chuneeta D. and {Patra}, Nipanjana and {Pieterse}, Samantha and {Pober}, Jonathan C. and {Razavi-Ghods}, Nima and {Ringuette}, Jon and {Robnett}, James and {Rosie}, Kathryn and {Sims}, Peter and {Smith}, Craig and {Syce}, Angelo and {Thyagarajan}, Nithyanandan and {Williams}, Peter K.~G. and {Zheng}, Haoxuan},
        title = "{Absolute Calibration Strategies for the Hydrogen Epoch of Reionization Array and Their Impact on the 21 cm Power Spectrum}",
      journal = {\apj},
         year = 2020,
        month = feb,
       volume = {890},
       number = {2},
          eid = {122},
        pages = {122},
          doi = {10.3847/1538-4357/ab67bc},
archivePrefix = {arXiv},
       eprint = {1910.12943},
 primaryClass = {astro-ph.IM},
       adsurl = {https://ui.adsabs.harvard.edu/abs/2020ApJ...890..122K}
}

@ARTICLE{Kern2021,
       author = {{Kern}, Nicholas S. and {Liu}, Adrian},
        title = "{Gaussian process foreground subtraction and power spectrum estimation for 21 cm cosmology}",
      journal = {\mnras},
         year = 2021,
        month = feb,
       volume = {501},
       number = {1},
        pages = {1463-1480},
          doi = {10.1093/mnras/staa3736},
archivePrefix = {arXiv},
       eprint = {2010.15892},
 primaryClass = {astro-ph.CO},
       adsurl = {https://ui.adsabs.harvard.edu/abs/2021MNRAS.501.1463K}
}

@ARTICLE{Kern2025,
       author = {{Kern}, Nicholas},
        title = "{A differentiable, end-to-end forward model for 21 cm cosmology: estimating the foreground, instrument, and signal joint posterior}",
      journal = {\mnras},
         year = 2025,
        month = aug,
       volume = {541},
       number = {2},
        pages = {687-713},
          doi = {10.1093/mnras/staf1007},
archivePrefix = {arXiv},
       eprint = {2504.07090},
 primaryClass = {astro-ph.CO},
       adsurl = {https://ui.adsabs.harvard.edu/abs/2025MNRAS.541..687K}
}

@inproceedings{SKALA4_ref1,
   title={Antenna design for the SKA1-LOW and HERA super radio telescopes},
   url={http://dx.doi.org/10.1109/ICEAA.2018.8520395},
   DOI={10.1109/iceaa.2018.8520395},
   booktitle={2018 International Conference on Electromagnetics in Advanced Applications (ICEAA)},
   publisher={IEEE},
   author={de Lera Acedo, Eloy and Pienaar, Hardie and Fagnoni, Nicolas},
   year={2018},
   month=sep, pages={636–639} }

@misc{SKALA4_ref2,
      title={SKA1-LOW Antenna Design Document}, 
      author={Eloy de Lera Acedo and Hardie Pienaar},
      year={2020},
      eprint={2003.12512},
      archivePrefix={arXiv},
      primaryClass={astro-ph.IM},
      url={https://arxiv.org/abs/2003.12512}, 
}

@ARTICLE{SKA_soil,
  author={Cavillot, Jean and Tihon, Denis and Gueuning, Quentin and de Lera Acedo, Eloy and Craeye, Christophe},
  journal={IEEE Transactions on Antennas and Propagation}, 
  title={Full-Wave Analysis of Thermal Noise in Antenna Arrays on Top of Layered Medium}, 
  year={2024},
  volume={72},
  number={10},
  pages={7560-7573},
  doi={10.1109/TAP.2024.3445132}}

@ARTICLE{SKA_soil_fastsim,
  author={Cavillot, Jean and Tihon, Denis and Mesa, Francisco and de Lera Acedo, Eloy and Craeye, Christophe},
  journal={IEEE Transactions on Antennas and Propagation}, 
  title={Efficient Simulation of Large Irregular Arrays on a Finite Ground Plane}, 
  year={2020},
  volume={68},
  number={4},
  pages={2753-2764},
  doi={10.1109/TAP.2019.2955180}}

@article{SKA_cables,
    author = {O’Hara, Oscar S D and Dulwich, Fred and de Lera Acedo, Eloy and Dhandha, Jiten and Gessey-Jones, Thomas and Anstey, Dominic and Fialkov, Anastasia},
    title = {Understanding spectral artefacts in SKA-Low 21-cm cosmology experiments: the impact of cable reflections},
    journal = {Monthly Notices of the Royal Astronomical Society},
    volume = {533},
    number = {3},
    pages = {2876-2892},
    year = {2024},
    month = {08},
    issn = {0035-8711},
    doi = {10.1093/mnras/stae1952},
    url = {https://doi.org/10.1093/mnras/stae1952},
    eprint = {https://academic.oup.com/mnras/article-pdf/533/3/2876/58975943/stae1952.pdf},
}

@misc{LFAA_station,
      title={SKA LFAA Station Design Report}, 
      author={Eloy de Lera Acedo and Hardie Pienaar and Nima Razavi Ghods and Jens Abraham and Edgar Colin Beltran and Ben Mort and Fred Dulwich and Giuseppe Virone and Benedetta Fiorelli and Michiel Arts and Christophe Craeye and Bui van Ha and Keith Grainge and Peter Dewdney and Jeff Wagg and Maria Grazia Labate and Andrew Faulkner and Jan Geralt bij de Vaate and Marchel Gerbers},
      year={2020},
      eprint={2003.12744},
      archivePrefix={arXiv},
      primaryClass={astro-ph.IM},
      url={https://arxiv.org/abs/2003.12744}, 
}

@article{SpectralII,
    author = {Trott, Cathryn M. and de Lera Acedo, Eloy and Wayth, Randall B. and Fagnoni, Nicolas and Sutinjo, Adrian T. and Wakley, Brett and Punzalan, Chris Ivan B.},
    title = {Spectral performance of Square Kilometre Array Antennas – II. Calibration performance},
    journal = {Monthly Notices of the Royal Astronomical Society},
    volume = {470},
    number = {1},
    pages = {455-465},
    year = {2017},
    month = {05},
    issn = {0035-8711},
    doi = {10.1093/mnras/stx1224},
    url = {https://doi.org/10.1093/mnras/stx1224},
    eprint = {https://academic.oup.com/mnras/article-pdf/470/1/455/17827642/stx1224.pdf},
}

@article{SKALA3,
    author = {de Lera Acedo, Eloy and Trott, Cathryn M. and Wayth, Randall B. and Fagnoni, Nicolas and Bernardi, Gianni and Wakley, Brett and Koopmans, Léon V.E. and Faulkner, Andrew J. and bij de Vaate, Jan Geralt},
    title = {Spectral performance of SKA Log-periodic Antennas I: mitigating spectral artefacts in SKA1-LOW 21 cm cosmology experiments},
    journal = {Monthly Notices of the Royal Astronomical Society},
    volume = {469},
    number = {3},
    pages = {2662-2671},
    year = {2017},
    month = {04},
    issn = {0035-8711},
    doi = {10.1093/mnras/stx904},
    url = {https://doi.org/10.1093/mnras/stx904},
    eprint = {https://academic.oup.com/mnras/article-pdf/469/3/2662/17628405/stx904.pdf},
}

@article{deleraacedo2015skala,
  title={SKALA, a log-periodic array antenna for the SKA-low instrument: design, simulations, tests and system considerations},
  author={de Lera Acedo, E. and Razavi-Ghods, N. and Troop, N. and Drought, N. and Faulkner, A. J.},
  journal={Experimental Astronomy},
  year={2015},
  volume={39},
  pages={567--594},
  doi={10.1007/s10686-015-9439-0},
  url={https://arxiv.org/abs/1512.01453}
}

@misc{ohara2025uncoupling,
  title={Uncovering the Effects of Array Mutual Coupling in 21-cm Experiments with the SKA-Low Radio Telescope},
  author={O'Hara, Oscar S. D. and Gueuning, Quentin and de Lera Acedo, Eloy and Dulwich, Fred and Cumner, John and Anstey, Dominic and Brown, Anthony and Fialkov, Anastasia and Dhandha, Jiten and Faulkner, Andrew and Liu, Yuchen},
  year={2025},
  archivePrefix={arXiv},
  primaryClass={astro-ph.CO},
  eprint={2412.01699},
  url={https://arxiv.org/abs/2412.01699}
}

@software{dulwich_2020_3758491,
  author       = {Dulwich, Fred},
  title        = {OSKAR 2.7.6},
  month        = jan,
  year         = 2020,
  publisher    = {Zenodo},
  version      = {2.7.6},
  doi          = {10.5281/zenodo.3758491},
  url          = {https://doi.org/10.5281/zenodo.3758491},
}

@INPROCEEDINGS{David,
  author={Gonzalez-Ovejero, David and Acedo, Eloy de Lera and Razavi-Ghods, Nima and Craeye, Christophe},
  booktitle={2009 IEEE Antennas and Propagation Society International Symposium}, 
  title={Fast MBF based method for large random array characterization}, 
  year={2009},
  volume={},
  number={},
  pages={1-4},
  doi={10.1109/APS.2009.5171749}}

@article{Bui_Van_2018,
   title={Fast and Accurate Simulation Technique for Large Irregular Arrays},
   volume={66},
   ISSN={1558-2221},
   url={http://dx.doi.org/10.1109/TAP.2018.2806222},
   DOI={10.1109/tap.2018.2806222},
   number={4},
   journal={IEEE Transactions on Antennas and Propagation},
   publisher={Institute of Electrical and Electronics Engineers (IEEE)},
   author={Bui-Van, Ha and Abraham, Jens and Arts, Michel and Gueuning, Quentin and Raucy, Christopher and Gonzalez-Ovejero, David and de Lera Acedo, Eloy and Craeye, Christophe},
   year={2018},
   month=apr, pages={1805–1817} }

@ARTICLE{Quentin2,
  author={Gueuning, Quentin and de Lera Acedo, Eloy and Keith Brown, Anthony and Craeye, Christophe and O’Hara, Oscar},
  journal={IEEE Transactions on Antennas and Propagation}, 
  title={A Broadband Multipole Method for Accelerated Mutual Coupling Analysis of Large Irregular Arrays Including Rotated Antennas}, 
  year={2025},
  volume={73},
  number={5},
  pages={3133-3145},
  doi={10.1109/TAP.2025.3528766}}

@ARTICLE{Quentin1,
  author={Gueuning, Quentin and de Lera Acedo, Eloy and Brown, Anthony Keith and Craeye, Christophe},
  journal={IEEE Transactions on Antennas and Propagation}, 
  title={An Inhomogeneous Plane-Wave Based Single-Level Fast Direct Solver for the Scattering Analysis of Extremely Large Antenna Arrays}, 
  year={2022},
  volume={70},
  number={10},
  pages={9511-9523},
  doi={10.1109/TAP.2022.3177465}}

@inbook{MutualCoupling,
author = {Lui, Hoi Shun Antony and Bird, Trevor S.},
publisher = {John Wiley \& Sons, Ltd},
isbn = {9781119565048},
title = {Mutual Coupling in Beamforming and Interferometric Antennas},
booktitle = {Mutual Coupling Between Antennas},
chapter = {10},
pages = {287-323},
doi = {https://doi.org/10.1002/9781119565048.ch10},
url = {https://onlinelibrary.wiley.com/doi/abs/10.1002/9781119565048.ch10},
eprint = {https://onlinelibrary.wiley.com/doi/pdf/10.1002/9781119565048.ch10},
year = {2021}
}

@INPROCEEDINGS{AnsteySKA,
  author={Anstey, Dominic and Cumner, John and Gueuning, Quentin and O'Hara, Oscar and de Lera Acedo, Eloy and Brown, Anthony and Faulkner, Andrew and Dulwich, Fred and Scott, Paul},
  booktitle={2024 18th European Conference on Antennas and Propagation (EuCAP)}, 
  title={Mitigating Zenith Blindness from Mutual Coupling in a Sunflower Phased Array}, 
  year={2024},
  volume={},
  number={},
  pages={1-5},
  doi={10.23919/EuCAP60739.2024.10501737}}

@article{HERAcoupling,
    author = {Josaitis, Alec T and Ewall-Wice, Aaron and Fagnoni, Nicolas and de Lera Acedo, Eloy},
    title = {Array element coupling in radio interferometry I: a semi-analytic approach},
    journal = {Monthly Notices of the Royal Astronomical Society},
    volume = {514},
    number = {2},
    pages = {1804-1827},
    year = {2022},
    month = {04},
    issn = {0035-8711},
    doi = {10.1093/mnras/stac916},
    url = {https://doi.org/10.1093/mnras/stac916},
    eprint = {https://academic.oup.com/mnras/article-pdf/514/2/1804/44055382/stac916.pdf},
}

@ARTICLE{SKALA41,
  author={Bolli, Pietro and Mezzadrelli, Lorenzo and Monari, Jader and Perini, Federico and Tibaldi, Alberto and Virone, Giuseppe and Bercigli, Mirko and Ciorba, Lorenzo and Di Ninni, Paola and Labate, Maria Grazia and Loi, Vittorio Giuseppe and Mattana, Andrea and Paonessa, Fabio and Rusticelli, Simone and Schiaffino, Marco},
  journal={IEEE Open Journal of Antennas and Propagation}, 
  title={Test-Driven Design of an Active Dual-Polarized Log-Periodic Antenna for the Square Kilometre Array}, 
  year={2020},
  volume={1},
  number={},
  pages={253-263},
  doi={10.1109/OJAP.2020.2999109}}

@INPROCEEDINGS{Quentin0,
  author={Gueuning, Q. and Craeye, C. and Colin-Beltran, E. and de Lera Acedo, E.},
  booktitle={2015 International Conference on Electromagnetics in Advanced Applications (ICEAA)}, 
  title={Validation of the HARP method for simulation of mutual coupling between SKALA antennas}, 
  year={2015},
  volume={},
  number={},
  pages={1214-1217},
  doi={10.1109/ICEAA.2015.7297311}}

@INPROCEEDINGS{cumner_cma,
  author={Cumner, John and O’Hara, Oscar S. D. and Gueuning, Quentin and Anstey, Dominic and Brown, Anthony and Dulwich, Fred and Faulkner, Andrew and Acedo, Eloy de Lera},
  booktitle={2025 International Conference on Electromagnetics in Advanced Applications (ICEAA)}, 
  title={Characteristic current mode analysis for an array of mutually coupled identical antennas}, 
  year={2025},
  volume={},
  number={},
  pages={1-4},
  note = {doi: Awaiting Publication}}

@INPROCEEDINGS{ohara_modelling,
  author={O'Hara, Oscar Sage David and Gueuning, Quentin and de Lera Acedo, Eloy and Dulwich, Fred and Anstey, Dominic and Cumner, John and Brown, Anthony and Faulkner, Andrew and Liu, Yuchen},
  booktitle={2025 19th European Conference on Antennas and Propagation (EuCAP)}, 
  title={Modelling Large Radio Telescope Visibilities Including Array Mutual Coupling Effects}, 
  year={2025},
  volume={},
  number={},
  pages={1-5},
  doi={10.23919/EuCAP63536.2025.10999651}}

@ARTICLE{Ali2008,
   author = {{Ali}, S.~S. and {Bharadwaj}, S. and {Chengalur}, J.~N.},
    title = "{Foregrounds for redshifted 21-cm studies of reionization: Giant Meter Wave Radio Telescope 153-MHz observations}",
  journal = {\mnras},
archivePrefix = "arXiv",
   eprint = {0801.2424},
     year = 2008,
    month = apr,
   volume = 385,
    pages = {2166-2174},
      doi = {10.1111/j.1365-2966.2008.12984.x},
   adsurl = {http://adsabs.harvard.edu/abs/2008MNRAS.385.2166A}
}

@ARTICLE{Bernardi2009,
   author = {{Bernardi}, G. and {de Bruyn}, A.~G. and {Brentjens}, M.~A. and 
	{Ciardi}, B. and {Harker}, G. and {Jeli{\'c}}, V. and {Koopmans}, L.~V.~E. and 
	{Labropoulos}, P. and {Offringa}, A. and {Pandey}, V.~N. and 
	{Schaye}, J. and {Thomas}, R.~M. and {Yatawatta}, S. and {Zaroubi}, S.
	},
    title = "{Foregrounds for observations of the cosmological 21 cm line. I. First Westerbork measurements of Galactic emission at 150 MHz in a low latitude field}",
  journal = {\aap},
archivePrefix = "arXiv",
   eprint = {0904.0404},
     year = 2009,
    month = jun,
   volume = 500,
    pages = {965-979},
      doi = {10.1051/0004-6361/200911627},
   adsurl = {http://adsabs.harvard.edu/abs/2009A%26A...500..965B}
}

@ARTICLE{Ghosh2012,
   author = {{Ghosh}, A. and {Prasad}, J. and {Bharadwaj}, S. and {Ali}, S.~S. and 
	{Chengalur}, J.~N.},
    title = "{Characterizing foreground for redshifted 21 cm radiation: 150 MHz Giant Metrewave Radio Telescope observations}",
  journal = {\mnras},
archivePrefix = "arXiv",
   eprint = {1208.1617},
 primaryClass = "astro-ph.CO",
     year = 2012,
    month = nov,
   volume = 426,
    pages = {3295-3314},
      doi = {10.1111/j.1365-2966.2012.21889.x},
   adsurl = {http://adsabs.harvard.edu/abs/2012MNRAS.426.3295G}
}

@ARTICLE{Paciga2013,
   author = {{Paciga}, G. and {Albert}, J.~G. and {Bandura}, K. and {Chang}, T.-C. and 
	{Gupta}, Y. and {Hirata}, C. and {Odegova}, J. and {Pen}, U.-L. and 
	{Peterson}, J.~B. and {Roy}, J. and {Shaw}, J.~R. and {Sigurdson}, K. and 
	{Voytek}, T.},
    title = "{A simulation-calibrated limit on the H I power spectrum from the GMRT Epoch of Reionization experiment}",
  journal = {\mnras},
archivePrefix = "arXiv",
   eprint = {1301.5906},
     year = 2013,
    month = jul,
   volume = 433,
    pages = {639-647},
      doi = {10.1093/mnras/stt753},
   adsurl = {http://adsabs.harvard.edu/abs/2013MNRAS.433..639P}
}

@article{Chatterjee2022,
    author = {Chatterjee, Suman and Bharadwaj, Somnath and Choudhuri, Samir and Sethi, Shiv and Patwa, Akash K},
    title = "{The tracking tapered gridded estimator for the power spectrum from drift scan observations}",
    journal = {\mnras},
    volume = {519},
    number = {2},
    pages = {2410-2425},
    year = {2022},
    month = {12},
    issn = {0035-8711},
    doi = {10.1093/mnras/stac3576},
    url = {https://doi.org/10.1093/mnras/stac3576},
    eprint = {https://academic.oup.com/mnras/article-pdf/519/2/2410/48480335/stac3576.pdf}
}

@article{Choudhuri2016b,
author = {Choudhuri, Samir and Bharadwaj, Somnath and Chatterjee, Suman and Ali, Sk. Saiyad and Roy, Nirupam and Ghosh, Abhik},
title = {The visibility-based tapered gridded estimator (TGE) for the redshifted 21-cm power spectrum},
journal = {\mnras},
volume = {463},
number = {4},
pages = {4093},
year = {2016},
doi = {10.1093/mnras/stw2254},
URL = { + http://dx.doi.org/10.1093/mnras/stw2254},
eprint = {/oup/backfile/Content_public/Journal/mnras/463/4/10.1093_mnras_stw2254/2/stw2254.pdf}
}

@ARTICLE{Choudhuri2014,
   author = {{Choudhuri}, S. and {Bharadwaj}, S. and {Ghosh}, A. and {Ali}, S.~S.
	},
    title = "{Visibility-based angular power spectrum estimation in low-frequency radio interferometric observations}",
  journal = {\mnras},
archivePrefix = "arXiv",
   eprint = {1409.7789},
     year = 2014,
    month = dec,
   volume = 445,
    pages = {4351-4365},
      doi = {10.1093/mnras/stu2027},
   adsurl = {http://adsabs.harvard.edu/abs/2014MNRAS.445.4351C}
}

@ARTICLE{Ghosh2011a,
   author = {{Ghosh}, A. and {Bharadwaj}, S. and {Ali}, S.~S. and {Chengalur}, J.~N.
	},
    title = "{GMRT observation towards detecting the post-reionization 21-cm signal}",
  journal = {\mnras},
archivePrefix = "arXiv",
   eprint = {1010.4489},
 primaryClass = "astro-ph.CO",
     year = 2011,
    month = mar,
   volume = 411,
    pages = {2426-2438},
      doi = {10.1111/j.1365-2966.2010.17853.x},
   adsurl = {http://adsabs.harvard.edu/abs/2011MNRAS.411.2426G}
}

@ARTICLE{Ghosh2011b,
   author = {{Ghosh}, A. and {Bharadwaj}, S. and {Ali}, S.~S. and {Chengalur}, J.~N.
	},
    title = "{Improved foreground removal in GMRT 610 MHz observations towards redshifted 21-cm tomography}",
  journal = {\mnras},
archivePrefix = "arXiv",
   eprint = {1108.3707},
 primaryClass = "astro-ph.CO",
     year = 2011,
    month = dec,
   volume = 418,
    pages = {2584-2589},
      doi = {10.1111/j.1365-2966.2011.19649.x},
   adsurl = {http://adsabs.harvard.edu/abs/2011MNRAS.418.2584G}
}

@ARTICLE{Pal22,
       author = {{Pal}, Srijita and {Elahi}, Kh Md Asif and {Bharadwaj}, Somnath and {Ali}, Sk Saiyad and {Choudhuri}, Samir and {Ghosh}, Abhik and {Chakraborty}, Arnab and {Datta}, Abhirup and {Roy}, Nirupam and {Choudhury}, Madhurima and {Dutta}, Prasun},
        title = "{Towards 21-cm intensity mapping at z = 2.28 with uGMRT using the tapered gridded estimator I: Foreground avoidance}",
      journal = {\mnras},
         year = 2022,
        month = oct,
       volume = {516},
       number = {2},
        pages = {2851-2863},
          doi = {10.1093/mnras/stac2419},
archivePrefix = {arXiv},
       eprint = {2208.11063},
 primaryClass = {astro-ph.CO},
       adsurl = {https://ui.adsabs.harvard.edu/abs/2022MNRAS.516.2851P}
}

@ARTICLE{Elahi2023,
       author = {{Elahi}, Kh Md Asif and {Bharadwaj}, Somnath and {Ghosh}, Abhik and {Pal}, Srijita and {Ali}, Sk Saiyad and {Choudhuri}, Samir and {Chakraborty}, Arnab and {Datta}, Abhirup and {Roy}, Nirupam and {Choudhury}, Madhurima and {Dutta}, Prasun},
        title = "{Towards 21-cm intensity mapping at z = 2.28 with uGMRT using the tapered gridded estimator - II. Cross-polarization power spectrum}",
      journal = {\mnras},
         year = 2023,
        month = apr,
       volume = {520},
       number = {2},
        pages = {2094-2108},
          doi = {10.1093/mnras/stad191},
archivePrefix = {arXiv},
       eprint = {2301.06677},
 primaryClass = {astro-ph.CO},
       adsurl = {https://ui.adsabs.harvard.edu/abs/2023MNRAS.520.2094E}
}

@ARTICLE{Morales2004,
   author = {{Morales}, M.~F. and {Hewitt}, J.},
    title = "{Toward Epoch of Reionization Measurements with Wide-Field Radio Observations}",
  journal = {\apj},
   eprint = {astro-ph/0312437},
     year = 2004,
    month = nov,
   volume = 615,
    pages = {7-18},
      doi = {10.1086/424437},
   adsurl = {http://adsabs.harvard.edu/abs/2004ApJ...615....7M}
}

@ARTICLE{Parsons2009,
       author = {{Parsons}, Aaron R. and {Backer}, Donald C.},
        title = "{Calibration of Low-Frequency, Wide-Field Radio Interferometers Using Delay/Delay-Rate Filtering}",
      journal = {\aj},
         year = 2009,
        month = jul,
       volume = {138},
       number = {1},
        pages = {219-226},
          doi = {10.1088/0004-6256/138/1/219},
archivePrefix = {arXiv},
       eprint = {0901.2575},
 primaryClass = {astro-ph.IM},
       adsurl = {https://ui.adsabs.harvard.edu/abs/2009AJ....138..219P}
}

@ARTICLE{Datta2007,
   author = {{Datta}, K.~K. and {Choudhury}, T.~R. and {Bharadwaj}, S.},
    title = "{The multifrequency angular power spectrum of the epoch of reionization 21-cm signal}",
  journal = {\mnras},
   eprint = {astro-ph/0605546},
     year = 2007,
    month = jun,
   volume = 378,
    pages = {119-128},
      doi = {10.1111/j.1365-2966.2007.11747.x},
   adsurl = {http://adsabs.harvard.edu/abs/2007MNRAS.378..119D}
}

@article{Mondal2018,
    author = {Mondal, Rajesh and Bharadwaj, Somnath and Datta, Kanan K.},
    title = "{Towards simulating and quantifying the light-cone EoR 21-cm signal}",
    journal = {\mnras},
    volume = {474},
    number = {1},
    pages = {1390-1397},
    year = {2017},
    month = {11},
    issn = {0035-8711},
    doi = {10.1093/mnras/stx2888},
    url = {https://doi.org/10.1093/mnras/stx2888},
    eprint = {https://academic.oup.com/mnras/article-pdf/474/1/1390/22367723/stx2888.pdf},
}

@ARTICLE{2019MNRAS.483.5694B,
       author = {{Bharadwaj}, Somnath and {Pal}, Srijita and {Choudhuri}, Samir and {Dutta}, Prasun},
        title = "{A Tapered Gridded Estimator (TGE) for the multifrequency angular power spectrum (MAPS) and the cosmological H I 21-cm power spectrum}",
      journal = {\mnras},
         year = 2019,
        month = mar,
       volume = {483},
       number = {4},
        pages = {5694-5700},
          doi = {10.1093/mnras/sty3501},
archivePrefix = {arXiv},
       eprint = {1812.08801},
 primaryClass = {astro-ph.CO},
       adsurl = {https://ui.adsabs.harvard.edu/abs/2019MNRAS.483.5694B}
}

@article{Paul2016,
	doi = {10.3847/1538-4357/833/2/213},
	url = {https://doi.org/10.3847%2F1538-4357%2F833%2F2%2F213},
	year = 2016,
	month = {dec},
	publisher = {American Astronomical Society},
	volume = {833},
	number = {2},
	pages = {213},
	author = {Sourabh Paul and Shiv K. Sethi and Miguel F. Morales and K. S. Dwarkanath and N. Udaya Shankar and Ravi Subrahmanyan and N. Barry and A. P. Beardsley and Judd D. Bowman and F. Briggs and P. Carroll and A. de Oliveira-Costa and Joshua S. Dillon and A. Ewall-Wice and L. Feng and L. J. Greenhill and B. M. Gaensler and B. J. Hazelton and J. N. Hewitt and N. Hurley-Walker and D. J. Jacobs and Han-Seek Kim and P. Kittiwisit and E. Lenc and J. Line and A. Loeb and B. McKinley and D. A. Mitchell and A. R. Neben and A. R. Offringa and B. Pindor and J. C. Pober and P. Procopio and J. Riding and I. S. Sullivan and M. Tegmark and Nithyanandan Thyagarajan and S. J. Tingay and C. M. Trott and R. B. Wayth and R. L. Webster and J. S. B. Wyithe and Roger Cappallo and M. Johnston-Hollitt and D. L. Kaplan and C. J. Lonsdale and S. R. McWhirter and E. Morgan and D. Oberoi and S. M. Ord and T. Prabu and K. S. Srivani and A. Williams and C. L. Williams},
	title = {{DELAY} {SPECTRUM} {WITH} {PHASE}-{TRACKING} {ARRAYS}: {EXTRACTING} {THE} H i {POWER} {SPECTRUM} {FROM} {THE} {EPOCH} {OF} {REIONIZATION}},
	journal = {\apj}
}

@article{Li2019,
	doi = {10.3847/1538-4357/ab55e4},
	url = {https://doi.org/10.3847%2F1538-4357%2Fab55e4},
	year = 2019,
	month = {dec},
	publisher = {American Astronomical Society},
	volume = {887},
	number = {2},
	pages = {141},
	author = {W. Li and J. C. Pober and N. Barry and B. J. Hazelton and M. F. Morales and C. M. Trott and A. Lanman and M. Wilensky and I. Sullivan and A. P. Beardsley and T. Booler and J. D. Bowman and R. Byrne and B. Crosse and D. Emrich and T. M. O. Franzen and K. Hasegawa and L. Horsley and M. Johnston-Hollitt and D. C. Jacobs and C. H. Jordan and R. C. Joseph and T. Kaneuji and D. L. Kaplan and D. Kenney and K. Kubota and J. Line and C. Lynch and B. McKinley and D. A. Mitchell and S. Murray and D. Pallot and B. Pindor and M. Rahimi and J. Riding and G. Sleap and K. Steele and K. Takahashi and S. J. Tingay and M. Walker and R. B. Wayth and R. L. Webster and A. Williams and C. Wu and J. S. B. Wyithe and S. Yoshiura and Q. Zheng},
	title = {First Season {MWA} Phase {II} Epoch of Reionization Power Spectrum Results at Redshift 7},
	journal = {ApJ}
}

@article{Trott2020,
    author = {Trott, Cathryn M and Jordan, C H and Midgley, S and Barry, N and Greig, B and Pindor, B and Cook, J H and Sleap, G and Tingay, S J and Ung, D and Hancock, P and Williams, A and Bowman, J and Byrne, R and Chokshi, A and Hazelton, B J and Hasegawa, K and Jacobs, D and Joseph, R C and Li, W and Line, J L B and Lynch, C and McKinley, B and Mitchell, D A and Morales, M F and Ouchi, M and Pober, J C and Rahimi, M and Takahashi, K and Wayth, R B and Webster, R L and Wilensky, M and Wyithe, J S B and Yoshiura, S and Zhang, Z and Zheng, Q},
    title = "{Deep multiredshift limits on Epoch of Reionization 21 cm power spectra from four seasons of Murchison Widefield Array observations}",
    journal = {\mnras},
    volume = {493},
    number = {4},
    pages = {4711-4727},
    year = {2020},
    month = {02},
    issn = {0035-8711},
    doi = {10.1093/mnras/staa414},
    url = {https://doi.org/10.1093/mnras/staa414},
    eprint = {https://academic.oup.com/mnras/article-pdf/493/4/4711/32927265/staa414.pdf},
}

@article{Elahi2025,
    author = {Elahi, Khandakar Md Asif and Bharadwaj, Somnath and Chatterjee, Suman and Sarkar, Shouvik and Choudhuri, Samir and Sethi, Shiv and Patwa, Akash Kumar},
    title = {The Tracking Tapered Gridded Estimator for the 21-cm power spectrum from MWA drift scan observations – II. The missing frequency channels},
    journal = {Monthly Notices of the Royal Astronomical Society},
    volume = {540},
    number = {3},
    pages = {2745-2761},
    year = {2025},
    month = {05},
    issn = {0035-8711},
    doi = {10.1093/mnras/staf896},
    url = {https://doi.org/10.1093/mnras/staf896},
    eprint = {https://academic.oup.com/mnras/article-pdf/540/3/2745/63411598/staf896.pdf},
}

@article{Patwa2021,
    author = {Patwa, Akash Kumar and Sethi, Shiv and Dwarakanath, K S},
    title = "{Extracting the 21 cm EoR signal using MWA drift scan data}",
    journal = {\mnras},
    volume = {504},
    number = {2},
    pages = {2062-2072},
    year = {2021},
    month = {04},
    issn = {0035-8711},
    doi = {10.1093/mnras/stab989},
    url = {https://doi.org/10.1093/mnras/stab989},
    eprint = {https://academic.oup.com/mnras/article-pdf/504/2/2062/37518347/stab989.pdf},
}

@ARTICLE{Bharadwaj2005,
   author = {{Bharadwaj}, S. and {Ali}, S.~S.},
    title = "{On using visibility correlations to probe the HI distribution from the dark ages to the present epoch - I. Formalism and the expected signal}",
  journal = {\mnras},
   eprint = {astro-ph/0406676},
     year = 2005,
    month = feb,
   volume = 356,
    pages = {1519-1528},
      doi = {10.1111/j.1365-2966.2004.08604.x},
   adsurl = {http://adsabs.harvard.edu/abs/2005MNRAS.356.1519B}
}

@article{sumanm2020,
    author = "Majumdar, Suman and Kamran, Mohd and Pritchard, Jonathan R. and Mondal, Rajesh and Mazumdar, Arindam and Bharadwaj, Somnath and Mellema, Garrelt",
    title = "{Redshifted 21-cm bispectrum \textendash{} I. Impact of the redshift space distortions on the signal from the Epoch of Reionization}",
    eprint = "2007.06584",
    archivePrefix = "arXiv",
    primaryClass = "astro-ph.CO",
    doi = "10.1093/mnras/staa3168",
    journal = "Mon. Not. Roy. Astron. Soc.",
    volume = "499",
    number = "4",
    pages = "5090--5106",
    year = "2020"
}

@ARTICLE{kamran2021,
       author = {{Kamran}, Mohd and {Ghara}, Raghunath and {Majumdar}, Suman and {Mondal}, Rajesh and {Mellema}, Garrelt and {Bharadwaj}, Somnath and {Pritchard}, Jonathan R. and {Iliev}, Ilian T.},
        title = "{Redshifted 21-cm bispectrum - II. Impact of the spin temperature fluctuations and redshift space distortions on the signal from the Cosmic Dawn}",
      journal = {\mnras},
         year = 2021,
        month = apr,
       volume = {502},
       number = {3},
        pages = {3800-3813},
          doi = {10.1093/mnras/stab216},
archivePrefix = {arXiv},
       eprint = {2012.11616},
 primaryClass = {astro-ph.CO},
       adsurl = {https://ui.adsabs.harvard.edu/abs/2021MNRAS.502.3800K}
}

@ARTICLE{Watkinson_2022,
       author = {{Watkinson}, Catherine A. and {Greig}, Bradley and {Mesinger}, Andrei},
        title = "{Epoch of reionization parameter estimation with the 21-cm bispectrum}",
      journal = {\mnras},
         year = 2022,
        month = mar,
       volume = {510},
       number = {3},
        pages = {3838-3848},
          doi = {10.1093/mnras/stab3706},
archivePrefix = {arXiv},
       eprint = {2102.02310},
 primaryClass = {astro-ph.CO},
       adsurl = {https://ui.adsabs.harvard.edu/abs/2022MNRAS.510.3838W}
}

@ARTICLE{gill_eormulti,
       author = {{Gill}, Sukhdeep Singh and {Pramanick}, Suman and {Bharadwaj}, Somnath and {Shaw}, Abinash Kumar and {Majumdar}, Suman},
        title = "{The monopole and quadrupole moments of the epoch of reionization (EoR) 21-cm bispectrum}",
      journal = {\mnras},
         year = 2024,
        month = jan,
       volume = {527},
       number = {1},
        pages = {1135-1140},
          doi = {10.1093/mnras/stad3273},
archivePrefix = {arXiv},
       eprint = {2310.15579},
 primaryClass = {astro-ph.CO},
       adsurl = {https://ui.adsabs.harvard.edu/abs/2024MNRAS.527.1135G}
}

@ARTICLE{Gill_2025_est,
       author = {{Gill}, Sukhdeep Singh and {Bharadwaj}, Somnath},
        title = "{A Visibility-based 21 cm Bispectrum Estimator for Radio-interferometric Data}",
      journal = {arXiv e-prints},
         year = 2025,
        month = jun,
          eid = {arXiv:2506.10526},
        pages = {arXiv:2506.10526},
          doi = {10.48550/arXiv.2506.10526},
archivePrefix = {arXiv},
       eprint = {2506.10526},
 primaryClass = {astro-ph.CO},
       adsurl = {https://ui.adsabs.harvard.edu/abs/2025arXiv250610526S}
}

@ARTICLE{Gill_2025,
       author = {{Gill}, Sukhdeep Singh and {Bharadwaj}, Somnath and {Elahi}, Khandakar Md Asif and {Sethi}, Shiv K. and {Patwa}, Akash Kumar},
        title = "{The Epoch of Reionization 21 cm Bispectrum at z = 8.2 from MWA Data. I. Foregrounds and Preliminary Upper Limits}",
      journal = {\apj},
         year = 2025,
        month = oct,
       volume = {993},
       number = {1},
        pages = {56},
          doi = {10.3847/1538-4357/ae0463},
archivePrefix = {arXiv},
       eprint = {2507.04964},
 primaryClass = {astro-ph.CO},
       adsurl = {https://ui.adsabs.harvard.edu/abs/2025arXiv250704964S}
}

@ARTICLE{Aguirre_HERA_e2eSims_2022,
       author = {{Aguirre}, James E. and {Murray}, Steven G. and {Pascua}, Robert and {Martinot}, Zachary E. and {Burba}, Jacob and {Dillon}, Joshua S. and {Jacobs}, Daniel C. and {Kern}, Nicholas S. and {Kittiwisit}, Piyanat and {Kolopanis}, Matthew and {Lanman}, Adam and {Liu}, Adrian and {Whitler}, Lily and {Abdurashidova}, Zara and {Alexander}, Paul and {Ali}, Zaki S. and {Balfour}, Yanga and {Beardsley}, Adam P. and {Bernardi}, Gianni and {Billings}, Tashalee S. and {Bowman}, Judd D. and {Bradley}, Richard F. and {Bull}, Philip and {Carey}, Steve and {Carilli}, Chris L. and {Cheng}, Carina and {DeBoer}, David R. and {Dexter}, Matt and {de Lera Acedo}, Eloy and {Ely}, John and {Ewall-Wice}, Aaron and {Fagnoni}, Nicolas and {Fritz}, Randall and {Furlanetto}, Steven R. and {Gale-Sides}, Kingsley and {Glendenning}, Brian and {Gorthi}, Deepthi and {Greig}, Bradley and {Grobbelaar}, Jasper and {Halday}, Ziyaad and {Hazelton}, Bryna J. and {Hewitt}, Jacqueline N. and {Hickish}, Jack and {Julius}, Austin and {Kerrigan}, Joshua and {Kohn}, Saul A. and {La Plante}, Paul and {Lekalake}, Telalo and {Lewis}, David and {MacMahon}, David and {Malan}, Lourence and {Malgas}, Cresshim and {Maree}, Matthys and {Matsetela}, Eunice and {Mesinger}, Andrei and {Molewa}, Mathakane and {Morales}, Miguel F. and {Mosiane}, Tshegofalang and {Neben}, Abraham R. and {Nikolic}, Bojan and {Parsons}, Aaron R. and {Patra}, Nipanjana and {Pieterse}, Samantha and {Pober}, Jonathan C. and {Razavi-Ghods}, Nima and {Ringuette}, Jon and {Robnett}, James and {Rosie}, Kathryn and {Santos}, Mario G. and {Sims}, Peter and {Singh}, Saurabh and {Smith}, Craig and {Syce}, Angelo and {Thyagarajan}, Nithyanandan and {Williams}, Peter K.~G. and {Zheng}, Haoxuan and {HERA Collaboration}},
        title = "{Validation of the HERA Phase I Epoch of Reionization 21 cm Power Spectrum Software Pipeline}",
      journal = {\apj},
         year = 2022,
        month = jan,
       volume = {924},
       number = {2},
          eid = {85},
        pages = {85},
          doi = {10.3847/1538-4357/ac32cd},
archivePrefix = {arXiv},
       eprint = {2104.09547},
 primaryClass = {astro-ph.IM},
       adsurl = {https://ui.adsabs.harvard.edu/abs/2022ApJ...924...85A}
}

@ARTICLE{Lynch2021,
	author = {{Lynch}, C.~R. and {Galvin}, T.~J. and {Line}, J.~L.~B. and {Jordan}, C.~H. and {Trott}, C.~M. and {Chege}, J.~K. and {McKinley}, B. and {Johnston-Hollitt}, M. and {Tingay}, S.~J.},
	title = "{The MWA long baseline Epoch of reionisation survey{\textemdash}I. Improved source catalogue for the EoR 0 field}",
	journal = {Publications of the Astronomical Society of Australia},
	year = 2021,
	month = nov,
	volume = {38},
	eid = {e057},
	pages = {e057},
	doi = {10.1017/pasa.2021.50},
	archivePrefix = {arXiv},
	eprint = {2110.08400},
	primaryClass = {astro-ph.CO},
	adsurl = {https://ui.adsabs.harvard.edu/abs/2021PASA...38...57L}
}

@ARTICLE{Tingay2013,
	author = {{Tingay}, S.~J. and {Goeke}, R. and {Bowman}, J.~D. and {Emrich}, D. and 
	{Ord}, S.~M. and {Mitchell}, D.~A. and {Morales}, M.~F. and 
	{Booler}, T. and {Crosse}, B. and {Wayth}, R.~B. and {Lonsdale}, C.~J. and 
	{Tremblay}, S. and {Pallot}, D. and {Colegate}, T. and {Wicenec}, A. and 
	{Kudryavtseva}, N. and {Arcus}, W. and {Barnes}, D. and {Bernardi}, G. and 
	{Briggs}, F. and {Burns}, S. and {Bunton}, J.~D. and {Cappallo}, R.~J. and 
	{Corey}, B.~E. and {Deshpande}, A. and {Desouza}, L. and {Gaensler}, B.~M. and 
	{Greenhill}, L.~J. and {Hall}, P.~J. and {Hazelton}, B.~J. and 
	{Herne}, D. and {Hewitt}, J.~N. and {Johnston-Hollitt}, M. and 
	{Kaplan}, D.~L. and {Kasper}, J.~C. and {Kincaid}, B.~B. and 
	{Koenig}, R. and {Kratzenberg}, E. and {Lynch}, M.~J. and {Mckinley}, B. and 
	{Mcwhirter}, S.~R. and {Morgan}, E. and {Oberoi}, D. and {Pathikulangara}, J. and 
	{Prabu}, T. and {Remillard}, R.~A. and {Rogers}, A.~E.~E. and 
	{Roshi}, A. and {Salah}, J.~E. and {Sault}, R.~J. and {Udaya-Shankar}, N. and 
	{Schlagenhaufer}, F. and {Srivani}, K.~S. and {Stevens}, J. and 
	{Subrahmanyan}, R. and {Waterson}, M. and {Webster}, R.~L. and 
	{Whitney}, A.~R. and {Williams}, A. and {Williams}, C.~L. and 
	{Wyithe}, J.~S.~B.},
	title = "{The Murchison Widefield Array: The Square Kilometre Array Precursor at Low Radio Frequencies}",
	journal = {Publications of the Astronomical Society of Australia},
	archivePrefix = "arXiv",
	eprint = {1206.6945},
	primaryClass = "astro-ph.IM",
	year = 2013,
	month = jan,
	volume = 30,
	eid = {e007},
	pages = {e007},
	doi = {10.1017/pasa.2012.007},
	adsurl = {http://adsabs.harvard.edu/abs/2013PASA...30....7T}
}

@ARTICLE{Wayth2018,
	author = {{Wayth}, Randall B. and {Tingay}, Steven J. and {Trott}, Cathryn M. and {Emrich}, David and {Johnston-Hollitt}, Melanie and {McKinley}, Ben and {Gaensler}, B.~M. and {Beardsley}, A.~P. and {Booler}, T. and {Crosse}, B. and {Franzen}, T.~M.~O. and {Horsley}, L. and {Kaplan}, D.~L. and {Kenney}, D. and {Morales}, M.~F. and {Pallot}, D. and {Sleap}, G. and {Steele}, K. and {Walker}, M. and {Williams}, A. and {Wu}, C. and {Cairns}, Iver. H. and {Filipovic}, M.~D. and {Johnston}, S. and {Murphy}, T. and {Quinn}, P. and {Staveley-Smith}, L. and {Webster}, R. and {Wyithe}, J.~S.~B.},
	title = "{The Phase II Murchison Widefield Array: Design overview}",
	journal = {Publications of the Astronomical Society of Australia},
	year = 2018,
	month = nov,
	volume = {35},
	eid = {e033},
	pages = {e033},
	doi = {10.1017/pasa.2018.37},
	archivePrefix = {arXiv},
	eprint = {1809.06466},
	primaryClass = {astro-ph.IM},
	adsurl = {https://ui.adsabs.harvard.edu/abs/2018PASA...35...33W}
}

@ARTICLE{Line2020,
       author = {{Line}, J.~L.~B. and {Mitchell}, D.~A. and {Pindor}, B. and {Riding}, J.~L. and {McKinley}, B. and {Webster}, R.~L. and {Trott}, C.~M. and {Hurley-Walker}, N. and {Offringa}, A.~R.},
        title = "{Modelling and peeling extended sources with shapelets: A Fornax A case study}",
      journal = {\pasa},
         year = 2020,
        month = jul,
       volume = {37},
          eid = {e027},
        pages = {e027},
          doi = {10.1017/pasa.2020.18},
archivePrefix = {arXiv},
       eprint = {2005.09316},
 primaryClass = {astro-ph.IM},
       adsurl = {https://ui.adsabs.harvard.edu/abs/2020PASA...37...27L}
}

@ARTICLE{Barry2019,
       author = {{Barry}, N. and {Beardsley}, A.~P. and {Byrne}, R. and {Hazelton}, B. and {Morales}, M.~F. and {Pober}, J.~C. and {Sullivan}, I.},
        title = "{The FHD/eppsilon Epoch of Reionisation power spectrum pipeline}",
      journal = {\pasa},
         year = 2019,
        month = jul,
       volume = {36},
          eid = {e026},
        pages = {e026},
          doi = {10.1017/pasa.2019.21},
archivePrefix = {arXiv},
       eprint = {1901.02980},
 primaryClass = {astro-ph.IM},
       adsurl = {https://ui.adsabs.harvard.edu/abs/2019PASA...36...26B}
}

@INPROCEEDINGS{Jordan2025,
  author={Jordan, Christopher H. and Null, Dev and Trott, Cathryn M. and LB Line, Jack and Chege, J. Kariuki and Lynch, Christene R. and Nunhokee, Chuneeta D. and Sleap, Greg and Wayth, Randall B.},
  booktitle={2025 URSI Asia-Pacific Radio Science Meeting (AP-RASC)}, 
  title={MWA_HYPERDRIVE: Next generation calibration software for the Murchison Widefield Array radio telescope}, 
  year={2025},
  volume={},
  number={},
  pages={1-4},
  doi={10.46620/URSIAPRASC25/LSCN1310}}

@ARTICLE{Sokolowski2017,
	author = {{Sokolowski}, M. and {Colegate}, T. and {Sutinjo}, A.~T. and {Ung}, D. and {Wayth}, R. and {Hurley-Walker}, N. and {Lenc}, E. and {Pindor}, B. and {Morgan}, J. and {Kaplan}, D.~L. and {Bell}, M.~E. and {Callingham}, J.~R. and {Dwarakanath}, K.~S. and {For}, Bi-Qing and {Gaensler}, B.~M. and {Hancock}, P.~J. and {Hindson}, L. and {Johnston-Hollitt}, M. and {Kapi{\'n}ska}, A.~D. and {McKinley}, B. and {Offringa}, A.~R. and {Procopio}, P. and {Staveley-Smith}, L. and {Wu}, C. and {Zheng}, Q.},
	title = "{Calibration and Stokes Imaging with Full Embedded Element Primary Beam Model for the Murchison Widefield Array}",
	journal = {Publications of the Astronomical Society of Australia},
	year = 2017,
	month = nov,
	volume = {34},
	eid = {e062},
	pages = {e062},
	doi = {10.1017/pasa.2017.54},
	archivePrefix = {arXiv},
	eprint = {1710.07478},
	primaryClass = {astro-ph.IM},
	adsurl = {https://ui.adsabs.harvard.edu/abs/2017PASA...34...62S}
}

@ARTICLE{Jaiden2021,
       author = {{Cook}, Jaiden H. and {Seymour}, Nicholas and {Sokolowski}, Marcin},
        title = "{A calibration and imaging strategy at 300 MHz with the Murchison Widefield Array (MWA)}",
      journal = {\pasa},
         year = 2021,
        month = dec,
       volume = {38},
          eid = {e063},
        pages = {e063},
          doi = {10.1017/pasa.2021.55},
       adsurl = {https://ui.adsabs.harvard.edu/abs/2021PASA...38...63C}
}

@article{Offringa2012,
	author = {{Offringa} and {van de Gronde, J. J.} and {Roerdink, J. B. T. M.}},
	title = {A morphological algorithm for improving radio-frequency interference detection},
	DOI= "10.1051/0004-6361/201118497",
	url= "https://doi.org/10.1051/0004-6361/201118497",
	journal = {A\&A},
	year = 2012,
	volume = 539,
	pages = "A95",
	month = "",
}

@ARTICLE{Offringa2015,
       author = {{Offringa}, A.~R. and {Wayth}, R.~B. and {Hurley-Walker}, N. and {Kaplan}, D.~L. and {Barry}, N. and {Beardsley}, A.~P. and {Bell}, M.~E. and {Bernardi}, G. and {Bowman}, J.~D. and {Briggs}, F. and {Callingham}, J.~R. and {Cappallo}, R.~J. and {Carroll}, P. and {Deshpande}, A.~A. and {Dillon}, J.~S. and {Dwarakanath}, K.~S. and {Ewall-Wice}, A. and {Feng}, L. and {For}, B. -Q. and {Gaensler}, B.~M. and {Greenhill}, L.~J. and {Hancock}, P. and {Hazelton}, B.~J. and {Hewitt}, J.~N. and {Hindson}, L. and {Jacobs}, D.~C. and {Johnston-Hollitt}, M. and {Kapi{\'n}ska}, A.~D. and {Kim}, H. -S. and {Kittiwisit}, P. and {Lenc}, E. and {Line}, J. and {Loeb}, A. and {Lonsdale}, C.~J. and {McKinley}, B. and {McWhirter}, S.~R. and {Mitchell}, D.~A. and {Morales}, M.~F. and {Morgan}, E. and {Morgan}, J. and {Neben}, A.~R. and {Oberoi}, D. and {Ord}, S.~M. and {Paul}, S. and {Pindor}, B. and {Pober}, J.~C. and {Prabu}, T. and {Procopio}, P. and {Riding}, J. and {Udaya Shankar}, N. and {Sethi}, S. and {Srivani}, K.~S. and {Staveley-Smith}, L. and {Subrahmanyan}, R. and {Sullivan}, I.~S. and {Tegmark}, M. and {Thyagarajan}, N. and {Tingay}, S.~J. and {Trott}, C.~M. and {Webster}, R.~L. and {Williams}, A. and {Williams}, C.~L. and {Wu}, C. and {Wyithe}, J.~S. and {Zheng}, Q.},
        title = "{The Low-Frequency Environment of the Murchison Widefield Array: Radio-Frequency Interference Analysis and Mitigation}",
      journal = {\pasa},
         year = 2015,
        month = mar,
       volume = {32},
          eid = {e008},
        pages = {e008},
          doi = {10.1017/pasa.2015.7},
archivePrefix = {arXiv},
       eprint = {1501.03946},
 primaryClass = {astro-ph.IM},
       adsurl = {https://ui.adsabs.harvard.edu/abs/2015PASA...32....8O}
}

@ARTICLE{Nunhokee2025,
       author = {{Nunhokee}, C.~D. and {Null}, D. and {Trott}, C.~M. and {Barry}, N. and {Qin}, Y. and {Wayth}, R.~B. and {Line}, J.~L.~B. and {Jordan}, C.~H. and {Pindor}, B. and {Cook}, J.~H. and {Bowman}, J. and {Chokshi}, A. and {Ducharme}, J. and {Elder}, K. and {Guo}, Q. and {Hazelton}, B. and {Hidayat}, W. and {Ito}, T. and {Jacobs}, D. and {Jong}, E. and {Kolopanis}, M. and {Kunicki}, T. and {Lilleskov}, E. and {Morales}, M.~F. and {Pober}, J.~C. and {Selvaraj}, A. and {Shi}, R. and {Takahashi}, K. and {Tingay}, S.~J. and {Webster}, R.~L. and {Yoshiura}, S. and {Zheng}, Q.},
        title = "{Limits on the 21 cm Power Spectrum at z = 6.5{\textendash}7.0 from Murchison Widefield Array Observations}",
      journal = {\apj},
         year = 2025,
        month = aug,
       volume = {989},
       number = {1},
          eid = {57},
        pages = {57},
          doi = {10.3847/1538-4357/adda45},
archivePrefix = {arXiv},
       eprint = {2505.09097},
 primaryClass = {astro-ph.CO},
       adsurl = {https://ui.adsabs.harvard.edu/abs/2025ApJ...989...57N}
}

@ARTICLE{Dillon2017,
       author = {{Dillon}, Joshua S. and {Parsons}, Aaron R.},
        title = "{Redundant Array Configurations for 21 cm Cosmology}",
      journal = {\apj},
         year = 2016,
        month = aug,
       volume = {826},
       number = {2},
          eid = {181},
        pages = {181},
          doi = {10.3847/0004-637X/826/2/181},
archivePrefix = {arXiv},
       eprint = {1602.06259},
 primaryClass = {astro-ph.IM},
       adsurl = {https://ui.adsabs.harvard.edu/abs/2016ApJ...826..181D}
}

\end{document}